\documentclass[aps,prb,twocolumn,floats]{revtex4-2}
\usepackage{amssymb}
\usepackage{amsbsy}
\usepackage{amsmath}
\usepackage{color}
\usepackage{float}
\usepackage{graphicx} 
\usepackage[colorlinks,linktocpage,bookmarks=false,citecolor=blue,linkcolor=red,urlcolor=blue]{hyperref}

\newcommand{\beq}{\begin{equation}}
\newcommand{\eeq}{\end{equation}}
\newcommand{\bea}{\begin{eqnarray}}
\newcommand{\eea}{\end{eqnarray}}

\definecolor{deepgreen}{RGB}{0,100,0}
\def\bpsi{\Psi}
\def\bphi{{\varphi}_\lambda}
\def\bA{\check{A}}

\def\bK{\check{K}^\lambda}
\def\bg{\check{g}_\lambda}
\def\bG{\check{G}}
\def\bGamma{\check{\Gamma}}
\def\bxi{{\xi}_\lambda}
\begin{document}

%\title{Subgap states and spin current in altermagnets with triplet pairing}
%\title{Index-theoretic route to the subgap Andreev bands and $4\pi$ Josephson effect}
\title{Index-theoretic route to the subgap Andreev bands and topological response in Josephson junctions}

\author{Sinchan Ghosh$^1$, Srinjoy Ghosh$^2$, Arijit Kundu$^2$, and K. Sengupta$^1$}

\affiliation{$^{(1)}$School of Physical Sciences, Indian
Association for the Cultivation of Science, 2A and 2B Raja S. C.
Mullick Road, Jadavpur 700032, India \\
$^{(2)}$ Department of Physics, Indian Institute of Technology, Kanpur, Uttar Pradesh, India.}

\date{\today}

\begin{abstract}
 We demonstrate that the subgap Andreev bound states in a transparent Josephson junction, comprising of either chiral or non-chiral superconductors, can be viewed as a consequence of the index theorem in supersymmetric quantum mechanics.  We provide an exact solution for these states starting from the Bogoliubov-de Gennes (BdG) equations describing quasiparticles in such junctions. We demonstrate that the dispersion of these subgap states depends only on the asymptotic properties of the pair-potential and not on its local spatial variation. Our study reveals the crucial distinction between junctions of non-chiral $p$-wave superconductors and those of $s$-wave or chiral superconductors by analyzing the wavefunction of their subgap bound states. We find a stable topological response leading to the well-known $4\pi$ periodic Josephson effect protected against weak disorder potential for the non-chiral $p$-wave junctions; no such protection is found for  junctions of $s$-wave or chiral superconductors. We supplement our analytic results with numerical computation of the Josephson currents in such junctions using exact numerical Green functions and starting from a lattice model of an itinerant altermagnet which is expected to host triplet $p$-wave superconductivity with equal-spin-pairing. We also discuss the implications of our results for Josephson junctions away from the transparent limit.  

\end{abstract}

\maketitle

\section{ Introduction}
\label{intro} 

Josephson effect, which describes flow of current through junctions of two superconductors, constitutes one of the central paradigms of superconductivity \cite{joseph1,josephrev}. These junctions usually have two superconductors separated by a long metallic or a short insulating layer; the latter class of junctions is termed as tunnel junctions \cite{likh1}. The low-energy properties of these tunnel junctions, which are typically characterized by a dimensionless barrier potential $Z$ that mimics the effect of the short insulating layer, can be understood from their localized subgap Andreev bound states; the energy dispersion of these states depends on the relative phase 
between the two superconductors and their pairing symmetry \cite{jos1}. The dispersion of such states becomes particularly simple in the Kulik-Omelyanchuk (KO) or the transparent limit for which $Z=0$ \cite{likh1}.

In recent years, the robustness of edge modes of topological $p$-wave superconductors have been of much interest for their potential quantum computation related application \cite{kit1,jos1}. These edge modes give rise to a $4\pi$-periodic current–phase relationship in a Josephson junction \cite{kit1,jos1}, which serves as an identifiable signature of their topological origin. In most discussions in the literature this effect is explained in terms of Majorana zero modes and fermion-parity protection, or equivalently via bulk topological invariants of the superconducting phase \cite{kit1,refparity}. Such an effect can also be directly computed using the Bogoliubov-de Gennes (BdG) equations assuming a step-function dependence of the superconducting pair-potentials near the junctions \cite{jos1}. However, such a derivation does not bring out the reason for robustness of the Andreev bound state dispersion against, for example, local spatial variation of the pair potential and/or weak disorder at the interface. 

The realization of triplet non-chiral $p$-wave superconductivity has been noted in quasi-one-dimensional (1D) Bechgaard salts \cite{tmtsfexp1,tmtsfexp2,ishi1,ks0,vmy0}. However, more recently, the advent of altermagnets has opened up a new avenue of study of quantum magnetism \cite{alt1,alt2,alt3,alt4,alt5,alt6,alt7,alt8,alt9,alt10,alt11,alt12}. A class of such altermagnets, namely itinerant altermagnets, lead to the realization of spin split electronic bands hosting spin polarized Fermi surface with zero net magnetization. Such altermagnets has also led to the possibility of realization of a triplet $p$-wave supercoducting state with equal-spin-pairing \cite{revsup,sudbo1,sudbo2}. 

Superconductivity in these altermagnets may be induced or due to spin-orbit coupling \cite{sup1,sup2,sup3}
or may arise, in some models, due to interaction of spin-split fermions with underlying magnons \cite{sudbo1}. For the latter case, in the regime where the spin-split bands 
are well separated in momentum space, Ref.\ \cite{sudbo2} predicted a possible realization of the equal-spin triplet $p$-wave superconducting state. The rationale behind
such a prediction comes from the fact that phase space for opposite spin singlet pairing gets drastically reduced due to the presence of momentum split energy bands of the altermagnets; 
consequently, equal-spin triplet pairing is energetically favored over singlet or opposite-spin triplet pairing \cite{revsup,sudbo1,sudbo2}. Transport in junctions of superconducting altermagnets have also been extensively studied in recent years \cite{sct1,sct2,sct3,sct4,sct5,sct6}. In most of these studies, the altermagnets have been
described by an effective single-band model with a spin-dependent hopping terms on a square lattice. In what follows, we shall use such a lattice model for numerical study of Josephson effect in such systems. 

In the present work we develop a complementary viewpoint for short tunnel junctions in the KO limit and obtain the following results. First, we show that the BdG equations for superconducting tunnel junctions in the KO limit can be cast in a supersymmetric form \cite{susyref1,vic1} irrespective of the pairing symmetry of the superconductors. In this formulation, it is possible to identify the existence of a superpotential which can be expressed in terms of the pair-potential of the superconductors. 

Second, we show that such a superpotential always changes sign as one traverses through the junction; this indicates the existence of bound states as a consequence of the Witten index theorem \cite{index1}. We show that these bounds states are precisely the subgap Andreev bound states which controls the low-energy properties of these junctions. Interestingly, our analysis shows that in contrast to the usual supersymmetric quantum systems where the bound states occur at $E=0$, the present system constitutes a somewhat rare case where they can occur at finite energy (with respect to the the Fermi energy which is set to zero). Their dispersion depends on the relative phase of the junction leading to formation of Andreev bands.

Third, our work brings out the central reason behind topological response of the $p$-wave superconductors and the corresponding $4\pi$ Josephson effect. Our analysis indicates that the structure of the wavefunction of the subgap bound states for $p$-wave junctions necessitates that the matrix elements of any charged impurity or barrier potential between them vanish. Thus such a potential can not lead to hybridization between these bound states which is the key to providing stability to the $4\pi$ Josephson effect; our analysis clearly indicates the topological protection in the sense that the lack of hybridization does not depend either on the microscopic form of the impurity or barrier potential or the details of the local spatial variation of the pair potential. We also show that such a lack of hybridization does not occur for either non-chiral $s$-wave or chiral superconductors that we analyze in this work; for these junctions, the Josephson current is expected to be $2\pi$ periodic. 

Finally, we back up our analytical results with those obtained using exact numerics. To this end, we turn to a microscopic lattice model of a 2D altermagnet with equal-spin triplet $p$-wave pairing. In the presence of such pairing, the superconducting altermagnets give rise to topological edge-modes, which are highly anisotropic for the two spins. Using a Green's function formalism we analyze the Josephson response, which shows a 4$\pi$ periodic current phase relationship. Our numerical results confirm the analytical prediction that the edge-modes, and their topological protection leading to the $4 \pi$ periodicity, remain robust against weak localized interfacial impurities. Moreover, we find that the $4\pi$ periodic behavior of the Josephson current persists in the strong barrier potential limit (known as the Ambegaonkar-Baratoff (AB) limit); this  confirms its independence of the barrier potential strength. 

The plan of the rest of the work is as follows. In Sec.\ \ref{ncs1} we present a detailed analysis of several chiral and non-chiral Josephson junctions in the KO limit and demonstrate the topological robustness of the $4\pi$ periodic Josephson current for non-chiral $p$-wave superconductors. This is followed by Sec.\ \ref{sec:josephson}
where we use exact numerics to study a junction of non-chiral $p$-wave superconductors starting from a lattice model that serves as a minimal model for a superconducting altermagnets. Finally, we discuss our main results and conclude in Sec.\ \ref{sec:diss}. Some details of relevant calculation have been charted out in the appendices.

\section{Analysis of Josephson junction} 
\label{ncs1}

 \begin{figure}
 \begin{center}
 \includegraphics[width=0.5\textwidth]{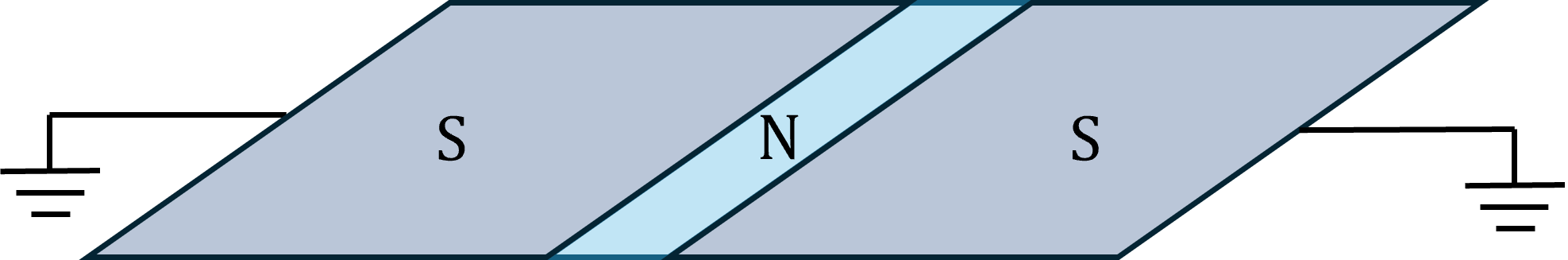}
 \caption{Schematic figure for the setup for a Josephson junction showing two superconductors (labeled as S) separated by a normal insulating region (labeled as N). The insulator is modeled by a potential $U_0 \gg t,\Delta_0$. The direction normal to the junction is taken to be $\hat x$ in all analysis throughout the work.} \label{figschss} 
 \end{center}	
 \end{figure}	

 In this section, we present an exact solution for Josephson junctions hosting chiral and non-chiral superconductors in the limit of transparent junctions. Our main point is to show that the Andreev bound states for such junctions can be understood, in this limit, as a consequence of an index theorem. 

The junction geometry is shown in Fig.\ \ref{figschss}. In what follows, we shall seek solutions for bound states in such junctions whose momenta is close to $k^F=|\vec k^F|$, where $\vec k^F= (k_x^F, \vec k^F_{\parallel})$. The direction normal to the junction is taken to be $\hat x$, and all transverse momenta $\vec k_{\parallel}$ are good quantum numbers. The BdG equations describing the right- and left-moving quasiparticles in the right and left superconductors in this regime can be written as \cite{jos1}
\begin{eqnarray} 
	E \left( \begin{array}{c} u_{\alpha \vec k_{\parallel}^F}^{\beta}(x) \\ v_{\alpha \vec k_{\parallel}^F}^{\beta}(x) \end{array} \right) &=& \left( \begin{array}{cc} \alpha \hat \zeta  & c_0 \Delta^{\beta}(x) \\ c_0 \Delta^{\beta \ast}(x) & -\alpha \hat \zeta \end{array} \right) 
	\left( \begin{array}{c} u_{\alpha \vec k_{\parallel}^F}^{\beta}(x) \\ v_{\alpha \vec k_{\parallel}^F}^{\beta}(x) \end{array} \right), \nonumber\\ \label{bdgjj1}
\end{eqnarray}
where $\hat \zeta= -i \hbar v_F(\vec k_{\parallel}^F) \partial_x = \hbar v_F(\vec k_{\parallel}^F) \hat k_x$,  $v_F$ denotes the Fermi velocity, $\alpha=\pm 1$ denotes right/left movers, $c_0=1(\alpha)$ for non-chiral $s(p)$-wave superconductors, and $\beta= R,L$ label the right and left superconductors. The choice of $c_0= \alpha$ for non-chiral $p$-wave superconductors is relevant for materials such as Bechgaard salts \cite{tmtsfexp1,tmtsfexp2} and altermagnets \cite{sudbo1} that have open quasi-1D Fermi surface; this will be discussed in Sec.\ \ref{sec:josephson}. For the chiral superconductors, $c_0= \exp[\pm i \varphi(k_{\parallel}^F)/2]$ where $\varphi(k_{\parallel}^F)$ is the momentum dependent phase of the pair-potential. The precise nature of $\varphi(k_{\parallel}^F)$ depends on the form of the pair-potential which shall be discussed in Sec.\ \ref{ssec3} in the context of chiral 2D superconductors. 

In this notation, the quasiparticle wavefunction is
\begin{eqnarray}
	\psi_{\alpha}^{\beta} (x,y) &=& e^{i (\alpha k^F_x x +\vec k_{\parallel}^F \cdot \vec r_{\parallel})}  \left( \begin{array}{c} u_{\alpha \vec k_{\parallel}^F}^{\beta}(x) \\ v_{\alpha \vec k_{\parallel}^F}^{\beta}(x) \end{array} \right). \label{wav1} 
\end{eqnarray} 
Here, and in the rest of this section, we set the global superconducting phase to zero without loss of generality. 

The Andreev bound state (ABS) energies for a generic short junction hosting non-chiral superconductors are known to be \cite{jos1}
	\begin{eqnarray} 
			E(\varphi) &=& \pm \Delta_0 \sqrt{D} \cos \varphi/2, \quad {\rm p-wave} \nonumber\\
			&=&  \Delta_0 \sqrt{1- D \sin^2\varphi/2}, \quad {\rm s-wave} \label{suben1}
		\end{eqnarray} 
where $\Delta_0$ is the gap amplitude, $\varphi$ the phase difference, and $D=4/(Z^2+4)$ is the transmission coefficient with the dimensionless barrier $Z= 2 m U_0/(\hbar k_F)$. The transparent or the KO limit is $Z=0$ for which $D=1$. We shall analyze the junction in this limit without assuming any specific spatial profile of the pair potential across the interface. Similar expressions has also been derived for 2D chiral superconductors in Ref.\ \onlinecite{jos1}; these depend on the nature of $\varphi(k_{\parallel}^F)$ and shall be discussed in Sec.\ \ref{ssec3}. 

 In what follows, we provide an explicit solution of the subgap Andreev bound states for junctions with non-chiral superconductors in Sec.\ \ref{ssec1}, discuss the stability of the $4\pi$ Josephson effect for $p$-wave in Sec.\ \ref{ssec2}, and chart out the solution for chiral superconductors in Sec.\ \ref{ssec3}. 

\subsection{ Solution for junctions of non-chiral superconductors}
\label{ssec1} 

 In this section, we carry out analysis of the subgap Andreev bound state hosting non-chiral $p$-wave and $s$-wave superconductors. In what follows, we show that the BdG equations can be expressed as a couple of equations that obey supersymmetric quantum mechanics and that the subgap Andrrev bound states stems from the associated Witten index theorem \cite{index1}. 

To this end, we first specify the asymptotic properties of the pair potential. Far from the junction, $\Delta(x=\pm \infty) = c_0 \Delta_0 e^{\pm i\varphi/2}$. Moreover, in the KO limit with $\varphi=0$, the junction behaves as a single superconductor so that $\Delta(x=0;\varphi=0)= c_0 \Delta_0$. For $0 \le \varphi\le 2\pi$, we can therefore choose $\Delta^{\beta}(x)\equiv c_0 \Delta(x)$ with 
\begin{eqnarray} 
	\Delta(x) &=& \Delta_1 + i\, \Delta_2\, f(\kappa_0 x), \quad \Delta_{1} = \Delta_0 \cos \frac{\varphi}{2}, \label{ppform1} \\ 
	\Delta_{2} &=&  \Delta_0 \sin \frac{\varphi}{2}= \sqrt{\Delta_0^2-\Delta_1^2},\qquad \kappa_0=\frac{\Delta_2}{\hbar v_F(\vec k_{\parallel}^F)}. \nonumber
\end{eqnarray} 
In what follows, we shall demand that $f(\pm \infty)=\pm 1$ and $f(0)=0$ so that $\Delta(x=\pm \infty) = \Delta_0 e^{\pm i\varphi/2}$
and $\Delta(x=0;\varphi=0)= \Delta_0$. Here $\kappa_0$ is the coherence length of the superconductors which depends on $\Delta_2$; we
note that the choice of $\kappa_0$ is motivated for capturing the correct asymptotic properties of $\Delta$ at $x=\pm \infty$. As we shall see, the specific spatial profile $f(x)$ is not important for the existence and dispersion of the subgap states; only its sign change across the junction matters.

	 In the transparent limit, there is no scattering between the right and the left moving quasiparticles at the junction and these branches can
     be treated separately.  First, we consider right movers and write $u_+(x)= u_+^R(x)$ for $x>0$ and $u_+^L(x)$ for $x<0$ (similarly for $v_+$), with continuity at $x=0$. Here and in the rest of this section, we shall suppress the transverse momentum dependence of $u$ and $v$ to avoid clutter. The BdG equations for the 
	right moving quasiparticles (with the chosen form of the pair-potential given by Eq.\ \ref{ppform1}), for which $c_0=1$ for both $s$- and $p$- wave, thus read
	\begin{eqnarray} 
		\hat \zeta u_+(x) + \Delta(x) v_+(x) &=& E u_+(x), \nonumber\\
		-\hat \zeta  v_+(x) + \Delta^{\ast}(x) u_+(x) &=& E v_+(x).   \label{bdgjj2}
	\end{eqnarray} 
	Defining $\psi_1(x) = u_+(x) + v_+(x)$ and $\psi_2(x)= u_+(x)-v_+(x)$, one can recast Eq.\ \ref{bdgjj2} as 
	\begin{eqnarray} 
		(\hat \zeta +i \Delta_2 f(\kappa_0 x)) \psi_1(x) &=& (E+\Delta_1) \psi_2(x), \nonumber\\
		(\hat \zeta -i \Delta_2 f(\kappa_0 x)) \psi_2(x) &=& (E-\Delta_1) \psi_1(x). \label{bdgjj3} 
	\end{eqnarray} 
	A straightforward decoupling of these equations yield
	\begin{eqnarray}
		H_{1(2)} \psi_{1(2)} &=& \epsilon^2 \psi_{1(2)}, \,\, H_1 = A_0 A_0^{\dagger}, \,\, H_2= A_0^{\dagger} A_0 \nonumber\\
		A_0 &=& (\hat \zeta -i \Delta_2 f(\kappa_0x)), \quad \epsilon^2 = E^2-\Delta_1^2   \label{bdgjj4}
	\end{eqnarray} 
	We note that $H_1$ and $H_2$ are supersymmetric partners of each other with the superpotential $\varphi(x) =  \Delta_2 f(\kappa_0 x)$. Furthermore, consistency with the asymptotic properties of the pair-potential discussed earlier mandates that $\varphi(x)$ must cross zero an odd number of times between $-\infty < x <\infty$. Thus one of these 
	Hamiltonian must host a solution $\epsilon=0$ or $E= \pm \Delta_1$, i.e, $E(\varphi) = \pm \Delta_0 \cos(\varphi/2)$. This shows that the subgap Andreev bound states are indeed a 
	consequence of the Witten index theorem \cite{index1}; their existence is guaranteed by the sign change of the superpotential (identified here with the imaginary part of the pair-potential) and is independent
	of the precise functional form of $\Delta(x)$. We note the present case forms a somewhat rare example where these bound states may have finite energy; they correspond to 
	$\epsilon=0$ (rather than $E=0$) and hence can span an energy band $-\Delta_0 \le E \le \Delta_0$ as $\varphi$ is varied. 
	
	Explicit solutions for these subgap states 
	can be easily obtained without assuming a specific form for $f(x)$. From Eq.\ \ref{bdgjj3}, we find that 
	the normalizable solutions for such Andreev bound states correspond to $E=-\Delta_1$ and $\psi_1=0$. This leads to, for both $p$- and $s$-wave junctions, 
	\begin{eqnarray} 
		E_1^{(p/s)} &=& -\Delta_0 \cos \varphi/2, \label{rsol1} \\
		\psi_{+}^{(p/s)}(x,y) &=& e^{i (k^F_x x +\vec k_{\parallel}^F \cdot \vec r_{\parallel})}  g(x) \left( \begin{array}{c} 1 \\ -1 \end{array} \right), \nonumber\\
        g(x) &=& N_0 \exp[- \kappa_0 \int^x dx' f(\kappa_0 x')], \nonumber
	\end{eqnarray}   
		where the normalization constant $N_0$ is determined from the condition $\int_{-\infty}^{\infty} dx |\psi(x)|^2 =1$.
	
	The other subgap branch can be obtained by studying the BdG equations for the left-movers. A simple analysis yields 
	\begin{eqnarray} 
		-\hat \zeta u_-(x) \pm \Delta(x) v_-(x) &=& E u_-(x) \nonumber\\
		\hat \zeta  v_-(x) \pm \Delta^{\ast}(x) u_-(x) &=& E v_-(x)   \label{bdgjj5}
	\end{eqnarray} 
	where the $+(-)$ sign corresponds to the $s(p)$-wave superconductors. We immediately note that for the $p$-wave junction, Eq.\ \ref{bdgjj5} identical to Eq.\ \ref{bdgjj2}
	with $E \to -E$. This allows us to write, for $p$-wave junctions,  the solution for the left-moving subgap branch  to be 
	\begin{eqnarray} 
		E_2^{(p)} &=&  \Delta_0 \cos \varphi/2, \nonumber \\
		\psi_{-}^{(p)}(x,y) &=& e^{i (- k^F_x x +\vec k_{\parallel}^F \cdot \vec r_{\parallel})}  g(x) \left( \begin{array}{c} 1 \\ -1 \end{array} \right). \label{rsol2} 
	\end{eqnarray} 
    
	The BdG equations for a $s$-wave junction do not have a similar property. From Eq.\ \ref{bdgjj5}, we find that they can be written as  
	\begin{eqnarray} 
		(\hat \zeta +i \Delta_2 f(\kappa_0 x)) \psi_2^{(s)}(x) &=& -(E - \Delta_1) \psi_1^{(s)}(x) \nonumber\\
		(\hat \zeta -i \Delta_2 f(\kappa_0 x)) \psi_1^{(s)}(x) &=& -(E +\Delta_1) \psi_2^{(s)}(x) \label{bdgsjj3} 
	\end{eqnarray} 
	These correspond to a set of supersymmetric Hamiltonians 
	\begin{eqnarray}
		H_{1(2)}^{(s)} \psi_{1(2)} &=& \epsilon_s^2 \psi_{1(2)}, \,\, H_1^{(s)} = A_0^{\dagger} A_0, \,\, H_2^{(s)}= A_0 A_0^{\dagger} \nonumber\\
		A_0 &=& (\hat \zeta -i \Delta_2 f(\kappa_0x)), \quad \epsilon_s^2 = E^2-\Delta_1^2 ,  \label{bdgsjj4}
	\end{eqnarray}
i.e, the partner ordering between $H_1$ and $H_2$ is swapped. The solutions for the subgap states, which once again must exist as a consequence of the index theorem, can be read off from Eq.\ \ref{bdgsjj3}. 
	The normalizable solution now corresponds to $E=\Delta_1$ and thus yields
	\begin{eqnarray} 
		E_2^{(s)} &=&  \Delta_0 \cos \varphi/2, \nonumber \\
		\psi_{-}^{(s)}(x,y) &=& e^{i (- k^F_x x +\vec k_{\parallel}^F \cdot \vec r_{\parallel})}  g(x) \left( \begin{array}{c} 1 \\ 1 \end{array} \right) \label{rsol3}
	\end{eqnarray} 

 Before ending this section, we note that a choice of specific $f(x)$ can also lead to solution of the extended eigenstates for which $E_n >\Delta_0$. This can be easily seen by choosing $f(\kappa_0 x)= \tanh(\kappa_0 x)$; such a choice allows an exact self-consistent solution of Eq.\ \eqref{bdgjj4} \cite{braz1}. With this choice one finds $H_1= \zeta^2+\Delta_0^2$ while the partner Hamiltonian $H_2$ has a Pöschl–Teller potential. For right movers and $E_n^2 > \Delta_0^2$ on a system of length $L$, one easily finds an explicit solution given by \cite{braz1}
\begin{eqnarray} 
	E_n^2 &=& (\hbar v_F n)^2 +\Delta_0^2, \quad v_F \equiv v_F(k_{\parallel}^F)\nonumber\\
	u_+(x) [v_+(x)] &=& \frac{e^{i n x}}{\sqrt{N(E_n) L}}\, c(x;E_n) \ [d(x;E_n)], \nonumber\\ 
	c(x;E_n) &=& \frac{E_n +  \Delta_1 + (\hbar v_F n +i  |\Delta_2| \tanh(\kappa_0 x))}{2(E_n +  \Delta_1)}, \nonumber\\
	d(x;E_n)  &=& \frac{(E_n + \Delta_1) -  (\hbar v_F n +i  |\Delta_2| \tanh(\kappa_0 x))}{2(E_n +\Delta_1)}, \nonumber\\
	N_1(E_n) &=&  \frac{ 2 E_n}{E_n + \Delta_1} \left(1- \frac{\hbar v_F n}{E_n (E_n+ \Delta_1) L} \right). \label{finenjj1}
\end{eqnarray}
A similar construction holds for left movers. The subgap energies remain $E=\pm \Delta_1$, as expected since they depend only on the sign change of $f(x)$; the ABS envelopes become $N_0 g(x) \to \sqrt{\kappa_0}\,{\rm sech}(\kappa_0 x)/2$, identifying $\kappa_0^{-1}$ as the localization length. We note that Eq.\ \ref{finenjj1} holds for the extended BdG quasiparticles near the Fermi surface in general; however, for quasi-1D superconducting junctions of superconducting Becchgard salts or altermagnets, where $v_F$ is only weakly dependent on the transverse momentum and the pair-potential is independent of it, they are expected to hold even for quasiparticles whose momenta are away from the Fermi surface.

\subsection{Stability of the $4\pi$ periodicity for $p$-wave junctions}
\label{ssec2} 

 Our analysis in the last section finds that the wavefunctions for the Andreev bound states originating from the left and the right movers for $p$-wave junctions have identical matrix structure. This indicates that the presence of a scatterer that couples to the charge $\hat n$ of these quasiparticles via  
 \begin{eqnarray} 
H_{\rm sc} = V_0 \int d^2 x\, g(\vec x) \hat n = V_0 \int d^2 x\,  g(\vec x)  \, \psi^{\dagger} \tau_3 \psi, \label{scpot1} 
\end{eqnarray} 
where $g(\vec x)$ provides the spatial profile of the potential centered around the junction, does not lead to a finite matrix element between these states since $\langle \psi_{1}^{(p)}|\tau_3|\psi_{2}^{(p)}\rangle=0$. This leads to the stability of $4\pi$ periodicity of such JJs against weak local perturbations.  We note that this statement is independent of the details of variation of the pair-potential across the junctions as well as the precise form of $g(\vec x)$. Such a stability is robust against both presence of non-magnetic impurities in the junction and a barrier potential $\sim U_0$; the latter feature explains the stability of $4\pi$ periodicity of the Josephson current for $p$-wave junctions for arbitrary barrier strength. 
	
This situation is to be contrasted with the subgap Andreev states for $s$-wave 
superconductors. For $s$-wave junctions, $\langle \psi_{+}^{(s)}|\tau_3|\psi_{-}^{(s)}\rangle \ne 0$ and thus a junction scatterer hybridizes the two subgap states leading to their $2 \pi$ periodicity. To see this explicitly, let us consider a potential localized at $x=0$ for which $g(\vec x)= \delta(x)$. In the space of the two states $|\pm\rangle$, where $\langle x|\pm \rangle = \psi_{\pm}^{(s)}$, we find that 
	\begin{eqnarray}
		\langle +|H_v|-\rangle &=& \langle -|H_v| +\rangle = V_1= V_0 [g(0)]^2.
	\end{eqnarray} 
	Thus in the space of the these two subgap states, one can write an effective Hamiltonian 
	\begin{eqnarray}
		H_{\rm eff} &=& \left(\begin{array}{cc} -\Delta_0\cos \varphi/2 & V_1 \\ V_1 & \Delta_0 \cos \varphi/2 \end{array} \right), \label{effham1}
	\end{eqnarray}
	A straightforward diagonalization of $H_{\rm eff}$ yields, after some algebra,
	\begin{align}
		E_{\pm}^{\rm eff} &=& \pm \Delta_0 \sqrt{T'} \sqrt{1- T' \sin^2\varphi/2}, ~ T'= \frac{1}{1+(V_1/\Delta_0)^2}. \label{effham2}
	\end{align}
	We note that Eq.\ \ref{effham2} yields subgap states energies which are $2\pi$ periodic and has the same form as that obtained from the standard scattering computation for SIS junctions \cite{jos1}. 
	
%	The contrasting situation between $p$-wave and $s$-wave junctions brings out the crucial role of the nature of the pair-potential for the stability of $4\pi$-periodicity of the subgap states 
%	in $p$-wave junctions. We note that we have not assumed any specific spatial form of the pair-potentials for demonstrating this stability. Our analysis relies only on the supersymmetric structure of the 
%	BdG equations. The existence of the subgap states and their waefunctions follows solely from the sign change of $f(x)$ across the junction; this stability can therefore be understood 
%	as a consequence of the index theorem. 
	
 Thus our analysis indicates that the stability of the $4\pi$ periodicity for  non-chiral $p$-wave junctions (and its fragility for $s$-wave junctions) can be understood as a consequence of the supersymmetric structure of the BdG equations and the corresponding index theorem. These dictate the independence of the Andreev bound states from local spatial variations of the pair-potential; moreover, they mandate the wavefunctions of these states  to obey a specific selection rule for the $\tau_3$ matrix element. This selection rule, in turn, leads to the absence ($p$-wave) or presence ($s$-wave) of hybridization of these bound states due to a barrier potential or non-magnetic impurities. 

For completeness, we briefly discuss the robustness of the 4$\pi$ periodic nature of the Josephson current phase relationship in the conventional Majorana representation at low energy in the App.\ \ref{app:majorana}. We note that the exact form of the wavefunctions of the Andreev modes depends on microscopic details of the Hamiltonian and can not be captured in this simplified Majorana based model, but the argument of the robustness of $4\pi$ periodic nature sustains.  The present analysis, however, indicates that the $4\pi$ periodic Josephson effect should remain stable in a much broader class of systems such as higher dimensional and spin-full triplet superconductors, where Majorana based models are inapplicable.

\subsection{Junctions of chiral Superconductors} 
\label{ssec3} 
	
The solution provided in Sec.\ \ref{ssec1} can be easily generalized to the case of transparent Josephson junctions between two chiral superconductors.  In what follows, we shall consider the left superconductor to have a pair-potential $\Delta_L(\vec k_{\parallel}^F) = \Delta_0 e^{-i(\varphi+\varphi(\vec k_{\parallel}^F))/2}$ far away from the junction. Here $\varphi(\vec k_{\parallel}^F)$ denotes the relative momentum-dependent phase of the pair-potential which depends on the Fermi momenta $\vec k_{\parallel}^F$. This phase typically differs for right and left moving quasiparticles; we shall denote $\varphi(\vec k_{\parallel}^F) =\varphi_{1(2)}$ for right(left) moving quasiparticles. For example, for a 2D chiral triplet $p$-wave superconductor with circular Fermi surface where $\Delta \sim (k_x^F-i \vec k_{y}^F)/k_F$, $k_{\parallel}^F= k_y^F$ and one has \cite{ruth1} 
	\begin{eqnarray} 
      \varphi(k_y^F) &=& \varphi_1 = 2 \sin^{-1}( k_y^F/k_F)  \quad  {\rm if }\, k_x^F >0 \nonumber\\
        &=&  \varphi_2= 2 ( \pi - \sin^{-1}( k_y^F/k_F)) \quad  {\rm if}\, k_x^F < 0 \label{leftsc}
	\end{eqnarray}
In what follows, unless explicitly mentioned, we shall not assume any specific form for $\varphi(\vec k_{\parallel}^F)$. For the right superconductor, we consider two possibilities.

\emph{Same chirality.} The right superconductor has the same chirality as the 
	left  so that far away to the right form the junction one has $\Delta_R(\vec k_{\parallel}^F) = \Delta_0 e^{i(\varphi - \varphi(\vec k_{\parallel}^F))/2}.$
	In this case, the momentum-dependent phase has the same sign structure in both superconductors. Following the steps of Sec.~\ref{ncs1}, one can write, for $0\le \varphi \le 2 \pi$,
	\begin{eqnarray} 
		\Delta(\vec k_{\parallel}^F;x) &=& e^{i \varphi(\vec k_{\parallel}^F)/2} \big(\Delta_1 + i \Delta_2 f(x)\big), \label{ppchiral1}
	\end{eqnarray}
	which differs from the non-chiral case only by an overall phase. The corresponding BdG equations for right/left movers are
	\begin{eqnarray} 
		\alpha \hat \zeta u_{\alpha} (x; \vec k_{\parallel}^F) +  \Delta(x; \vec k_{\parallel}^F) v_{\alpha}(x,\vec k_{\parallel}^F) &=& E u_{\alpha} (x; \vec k_{\parallel}^F), \nonumber\\
		- \alpha \hat \zeta v_{\alpha} (x; \vec k_{\parallel}^F) +  \Delta^{\ast}(x;\vec k_{\parallel}^F) u_{\alpha}(x,\vec k_{\parallel}^F) &=& E v_{\alpha} (x; \vec k_{\parallel}^F). \nonumber\\ \label{bdgchiral1}
	\end{eqnarray}
where $\alpha=1(-1)$ for the right(left) moving quasiparticles and $k_{\alpha x}^F= \alpha \sqrt{(k^F)^2-|\vec k_{\parallel}^F|^2 }$. 
Defining, for the right moving qusdiparticles, 
\begin{eqnarray} 
	\tilde u_{+} (x; \vec k_{\parallel}^F) &=&  e^{-i \varphi_1/4} u_{+}(x; \vec k_{\parallel}^F), \nonumber\\
	\tilde v_{+} (x; \vec k_{\parallel}^F) &=& e^{i \varphi_1/4} v_{+}(x;\vec k_{\parallel}^F), \label{varchiraldef}
\end{eqnarray}
Eq.\ \eqref{bdgchiral1} reduces to
\begin{eqnarray} 
	\hat \zeta \tilde u_{+}  +  (\Delta_1 + i \Delta_2 f(x)) \tilde v_{+}  &=& E \tilde u_{+},  \nonumber\\
	-\hat \zeta \tilde v_{+} + (\Delta_1 - i \Delta_2  f(x)) \tilde u_{+} &=& E \tilde v_{+}.  \label{bdgchiral2}
\end{eqnarray}
This is exactly the same equation obtained in Sec.\ \ref{ssec1} (Eqs.\ \ref{bdgjj3} and \ref{bdgsjj3}) for non-chiral superconductors and yields identical expressions of subgap energy states:
\begin{eqnarray} 
	E_{1} [\varphi] &=& -\Delta_0 \cos \frac{\varphi}{2}, \label{bdgchiral3}\\
	\psi_{1}(x,\vec k_{\parallel}^F) &=& e^{i (k_x^F x + \vec k_{\parallel}^F \cdot \vec r_{\parallel})} g(x) 
	\left(\begin{array}{c} e^{i \varphi_1/4} \\  -e^{-i \varphi_1/4} \end{array} \right) 
\end{eqnarray} 
where $g(x)= N_0 \exp[-\kappa_0 \int^x f(x') dx']$, $N_0$ is the normalization constant,  and $\kappa_0 = \Delta_2/(\hbar v_F(\vec k_{\parallel}^F))$. 

A similar analysis can be carried out for the left-moving quasiparticles. Here we define 
\begin{eqnarray} 
	\tilde u_{-} (x; \vec k_{\parallel}^F) &=&  e^{i \varphi_2/4} u_{-}(x; \vec k_{\parallel}^F), \nonumber\\
	\tilde v_{-} (x; \vec k_{\parallel}^F) &=& e^{-i \varphi_2/4} v_{-}(x;\vec k_{\parallel}^F), \label{varchiraldefa}
\end{eqnarray}
which leads to the equation 
\begin{eqnarray} 
	-\hat \zeta \tilde u_{-}  +  (\Delta_1 + i \Delta_2 f(x)) \tilde v_{-}  &=& E \tilde u_{-},  \nonumber\\
	+\hat \zeta \tilde v_{-} + (\Delta_1 - i \Delta_2  f(x)) \tilde u_{-} &=& E \tilde v_{-}.  \label{bdgchiral2b}
\end{eqnarray}
From the analysis of the $s$-wave non-chiral junctions, we can write down the solutions for the subgap energy states as 
identical expressions of subgap energy states:
\begin{eqnarray} 
	E_{2} [\varphi] &=& \Delta_0 \cos \frac{\varphi}{2}, \label{bdgchiral3}\\
	\psi_{2}(x,\vec k_{\parallel}^F) &=& e^{i (-k_x^F x + \vec k_{\parallel}^F \cdot \vec r_{\parallel})} g(x) 
	\left(\begin{array}{c} e^{i \varphi_2/4} \\  e^{-i \varphi_2/4} \end{array} \right) 
\end{eqnarray}

We note that the presence of a charged impurity scatterer with a potential given by Eq.\ \ref{scpot1} leads to a matrix element $\langle \psi_2|H_{\rm sc}|\psi_1\rangle \sim \cos (\varphi_2-\varphi_1)/4$. For the 2D triplet $p$-wave junction with a circular Fermi surface this yields 
\begin{eqnarray} 
\langle \psi_2|H_{\rm sc}|\psi_1\rangle \sim \cos(\pi/2 -\sin^{-1}(k_{y}^F/k_F)) \sim k_{y}^F/k_F. \label{chiralpsc}
\end{eqnarray}
Thus the presence of chirality leads to hybridization of the subgap bound states leading to $2\pi$ periodic Josephson current. Our result here reduces to that for non-chiral $p$-wave superconductor for which 
$\varphi_1=0$ and $\varphi_2=2\pi$; the scattering matrix elements vanish in this case leading to the $4\pi$ periodic Josephson effect.  

\emph{Opposite chirality.} Here we consider the right superconductor to have the \emph{opposite} chirality so that far away from the junction
\begin{eqnarray} 
	\Delta_R(\vec k_{\parallel}^F) &=& \Delta_0 e^{i(\varphi + \varphi(\vec k_{\parallel}^F) )/2}. \label{rightsc1}
\end{eqnarray}
	We note that in this situation, the momentum dependent chiral phase is no longer a global phase since it has opposite signs in the two superconductors. To analyze such a junction, we first consider the form of the pair potential given by 
\begin{eqnarray} 
	\Delta(\vec k_{\parallel}^F;x) &=& \Delta_1 + i \Delta_2 f(x), \nonumber\\
	\Delta_1 &=& \Delta_0 \cos \frac{\varphi + \varphi(\vec k_{\parallel}^F)}{2}, \label{ppchiral2}\\
	\Delta_2 &=& \Delta_0 \sin \frac{\varphi + \varphi(\vec k_{\parallel}^F)}{2}, \nonumber
\end{eqnarray} 
where the range of the relative phase is to be chosen so that $\kappa_1 \sim \Delta_2/\hbar v_F(k_y^F) \ge 0$ and we choose $f(\pm \infty)=\pm 1$ as before.
The BdG equations for superconducing quasiparticles can then be written as 
\begin{widetext} 
	\begin{eqnarray} 
		\alpha \hat \zeta u_{\alpha} (x; \vec k_{\parallel}^F) +  (\Delta_1 + i \Delta_2 f(x)) v_{\alpha} (x; \vec k_{\parallel}^F)  &=& E u_{\alpha} (x; \vec k_{\parallel}^F), \nonumber\\
		- \alpha \hat \zeta v_{\alpha} (x; \vec k_{\parallel}^F) + (\Delta_1 - i \Delta_2 f(x) ) u_{\alpha}(x,\vec k_{\parallel}^F) &=& E v_{\alpha} (x; \vec k_{\parallel}^F). \label{bdgchiral4}
	\end{eqnarray}
\end{widetext} 
Carrying out a similar analysis as in Sec.\ \ref{ssec1}, one finds 
	\begin{eqnarray} 
		E_{1(2)} &=& -(+) \Delta_0 \cos \frac{+(-)\varphi + \varphi_{1(2)}(\vec k_{\parallel}^F)}{2}  \label{bdgchiral5}\\
		\psi_{1(2)} &=&  e^{i (+(-) k_x^F x +  \vec k_{\parallel}^F \cdot \vec r_{\parallel})} g_1(x; \vec k_{\parallel}^F) \left(\begin{array}{c} 1 \\ -(+) 1 \end{array} \right), \nonumber\\
        g_1(x) &=& N_1 \exp[- \kappa_1 \int^x dx' f(\kappa_1 x')],  \nonumber
	\end{eqnarray} 
where $\alpha=\pm$ for the right and the left movers, $\kappa_1= \Delta_0 |\sin(\varphi_0 + \varphi(\vec k_{\parallel}^F))|/(\hbar v_F(\vec k_{\parallel}))$, and $N_1$ is obtained from the normalization of $\psi_{\pm}$. The solution is valid for range of $\phi$ for which $\kappa_1>0$. For the 2D triplet chiral $p$-wave supercoductors where $\varphi(\vec k_{\parallel}^F) \equiv \varphi(k_y^F)$ is given by Eq.\ \ref{leftsc}, we obtain for the right movers
\begin{eqnarray} 
&& E_{1} = -\Delta_0 \cos[\varphi/2 + \sin^{-1}(k_{y}^F/k_F)] \label{pwavchi1} \\
&=& \Delta_0 \Big[- \cos \frac{\varphi}{2} \sqrt{1-(k_{y}^F/k^F)^2} + (k_{y}^F/k^F) \sin \frac{\varphi}{2}\Big], \nonumber
\end{eqnarray} 
where we have used the fact that $k_x^F/k_F= \alpha \sqrt{1-(k_{y}^F/k_F)^2}/k_F$. A similar analysis for the left movers yield 
\begin{eqnarray} 
E_{2} &=&  \Delta_0 \Big[\cos \frac{\varphi}{2} \sqrt{1-(k_{y}^F/k^F)^2} + (k_{y}^F/k^F) \sin \frac{\varphi}{2}\Big]. \label{pwavchi2}
\end{eqnarray}
These subgap states can once again hybridize in the presence of impurities as is evident from their wavefunctions and 
hence we can expect $2 \pi$ periodic Josephson effect in a junction with finite barrier potential. We note that the expressions of the subgap energies obtained in Eqs.\ \ref{pwavchi1} and \ref{pwavchi2} 
matches with those found in Ref.\ \onlinecite{jos1} assuming a step function dependence of the pair-potential in the KO limit.

\section{Numerical results}
\label{sec:josephson}

 In this section, we back up our analytic results with numerical studies of $p$-wave non-chiral superconductors. Such superconductivity is known to occur in Bechgaard salts \cite{tmtsfexp1,tmtsfexp2}; more recently, they have also been proposed as superconducting phase in quantum altermagnets \cite{sudbo1}. In what follows, we shall use a model describing superconducting phase of quantum altermagnets for our numerical studies. The details of model is described in Sec.\ \ref{sec:alt} while our numerical results on Josephson junctions of such altermagents is presented in Sec.\ \ref{sec:jr}. 

\subsection{Altermagnets}
\label{sec:alt}

Altermagnets have up and down spin Fermi surfaces that rotate in momentum space, keeping the SU(2) symmetry of the underlying Hamiltonian intact. This unique magnetic nature of the bands led to exploration of interaction driven effects, including possible superconductivity. 
Recently it has been proposed that intrinsic superconductivity in altermagnets can be of $p$-wave type  with equal-spin pairing\cite{sudbo1,sudbo2}. This immediately leads to the question of topological transport signatures in a Josephson setup, and whether these signatures are protected against small impurities. In this section we present our study of Josephson responses of altermagnets, and their robustness.

%---------------------------
\begin{figure}[!t]
	\begin{center}
		\includegraphics[width=1.0\linewidth]{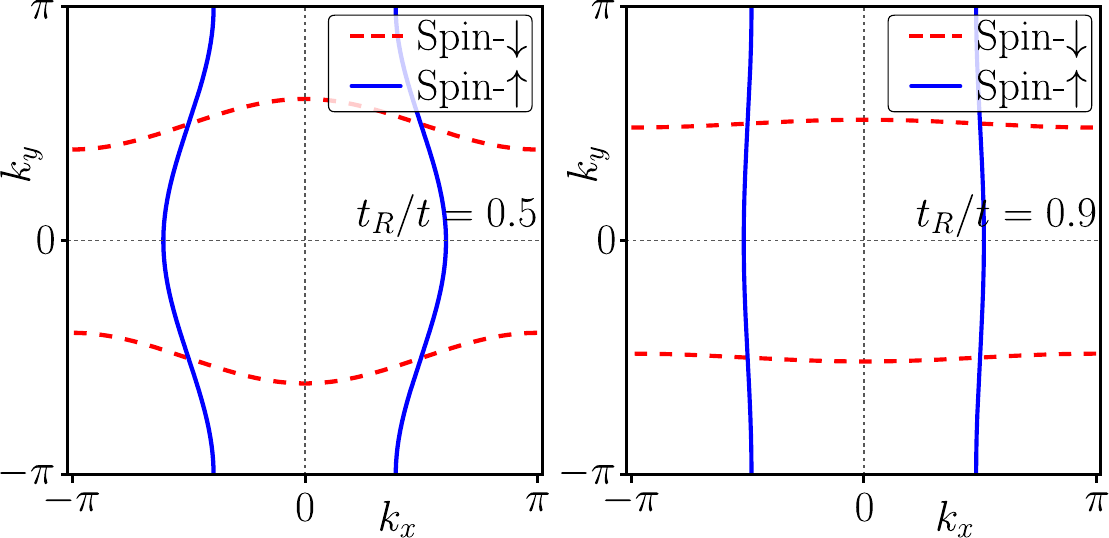}
		\caption{Plot of the Fermi surface of an altermagnets obtained from Eq.\ \eqref{disp} for chemical potential $\mu / t = 0$ and $\eta=t_R /t=0.5$ (left panel) and $0.9$ (right panel). The Fermi lines flatten out as $\eta$ approaches unity. The spin labels, shown in the figures, indicates anisotropic spin-momentum locking.} \label{fs} 
	\end{center}	
\end{figure}
%----------------------------

Following recent proposals~\cite{sct1,sct2,sct3,sct4,sct5,sct6,sudbo1,sudbo2}, we consider a simple two band (one for each of the spin sector) Hamiltonian for altermagnets:
\begin{eqnarray}
H &=& -\sum_{\vec j} \left[  c_{\sigma}^{\dagger} (j_x,j_y) \Big[(t+{\rm sgn}(\sigma) t_R ) c_{\sigma}(j_x+1,j_y) \right. \nonumber \\
&& \left. + (t-{\rm sgn}(\sigma) t_R) c_{\sigma}(j_x,j_y+1)  \Big] +{\rm  h.c.} \right]\nonumber\\
&& -\mu \sum_j c_{\sigma}^{\dagger}(j_x,j_y)  c_{\sigma} (j_x,j_y),  \label{ham1}
\end{eqnarray} 
where $c_{\sigma}(j_x,j_y)$ denotes annihilation operator of a fermion with spin $\sigma =\uparrow,\downarrow$ on the site $\vec j=(j_x,j_y)$ 
of a 2D square lattice, $t$ is the amplitude of fermion hopping, $t_R$ denotes the amplitude of spin-dependent hopping which takes opposite values for spin-up and spin-down fermions, $\mu$ is the chemical potential, and ${\rm sgn}(\sigma) = 1(-1)$ for $\sigma=\uparrow(\downarrow)$.
The energy dispersion of such fermions are given by
\begin{align} 
\epsilon_{k \sigma} &= -2t \left[ (\cos k_x+ \cos  k_{y}) \nonumber \right. \\
&+ \left. \eta (\cos k_x-\cos  k_{y}) {\rm sgn}(\sigma) \right] - \mu, 
\label{disp}
\end{align}  
where $\vec k=(k_x, k_{y})$ denotes quasimomentum and $\eta=t_R/t$. The spin-split Fermi surface for chemical potential $\mu / t = 0$ is shown in Fig.\ \ref{fs}, for $\eta=0.5$ and $0.9$. We notice that the curvature of the Fermi surfaces reduce as the value of $\eta$ increases; they become parallel to $k_x$ (spin-up) and $k_y$ (spin-down) as $\eta$ approaches unity. In this limit, spin-up (spin-down) fermions predominatly disperse along $k_x(k_y)$.

%------------------------------
%\begin{figure*}[ht!bp]
% 	\begin{center}
% 		\includegraphics[width=1.02\textwidth]{FinalPlots/abs.pdf}
% \caption{Andreev bound state spectrum of the Josephson junction between two effective 1D superconductors with fixed $\vec k_{\parallel} = 0$, with a weak-link hopping $t_J$, for a few values of $t_R / t$. The number of sites of each superconductor is $N=120$ along the $x$-direction, and $\Delta_0 / t = 0.1$, $\mu / t = 0$ and $t_J / t = 0.01$. The subgap energy spectrum versus $\varphi$ shows a gap of $2\delta$ at $\varphi = \pi$, where $\delta$ in the energy splitting between Majorana edge modes in each superconductor in the absence of the junction ($t_J / t = 0$). We note that the gap $2\delta$ increases exponentially with increase in $t_R / t$. \label{abs} }
% 	\end{center}	
% \end{figure*}
%--------------------------------

In order to model superconductivity in an altermagnetic metal, we assume an equal-spin triplet pairing. The equal-spin pairing of fermions in these systems can arise due to their interaction with magnons in certain models; the details of such pairing have been studied in Refs.\ \onlinecite{sudbo1,sudbo2}. Such an equal-spin triplet pairing is given by the pairing potential
\begin{eqnarray}
\Delta_{\uparrow \uparrow}(\vec k) &=& \langle c_{\uparrow}(\vec k) c_{\uparrow}(-\vec k)\rangle = \Delta_0 \frac{k_{x}^{F}}{|{\vec k_F}|}  \nonumber\\
\Delta_{\downarrow \downarrow}(\vec k) &=& \langle c_{\downarrow}(\vec k) c_{\downarrow}(-\vec k)\rangle = \Delta_0 \frac{k_{y}^{F}}{|{\vec k_F}|} 
\label{pair_p}
\end{eqnarray}
where ${\vec k_F}$ denotes the Fermi momentum, $k_x^{F}$ and $ k_{y}^{F}$ are its $x$ and $y$ component respectively such that $|{\vec k_F}|=\sqrt{(k_{x}^{F})^2+(k_{y}^{F})^2}$. Note that the pairing is odd under exchange of momentum and that spin-$\uparrow$ ($\downarrow$) fermions pair with relative momentum aligned along the $x$($y$)-direction, consistent with the anisotropic band structure of the altermagnet.

For the lattice calculations and numerics that follow, we consider a microscopically consistent real-space pairing Hamiltonian which captures the same symmetry structure as the pair potentials above. We introduce nearest-neighbour equal-spin pairing on a 2D square lattice with the pairing bonds oriented along the direction in which each spin carrier predominantly propagates. The pairing Hamiltonian is given by
\begin{eqnarray}
    H_{\text{pair}} &=& \sum_{\vec j,\sigma} \left[ \Delta_0 e^{i \phi} \,c^\dag_{\uparrow}(j_x + 1, j_y)c^\dag_{\uparrow}(j_x, j_y) \right. \nonumber \\
    && \left. + \Delta_0 e^{i \phi} \,c^\dag_{\downarrow}(j_x, j_y + 1)c^\dag_{\downarrow}(j_x, j_y)+\mathrm{h.c.} \right],
\label{hpair}
\end{eqnarray}
where $\Delta_0$ is the $p$-wave pairing amplitude and $\phi$ is the superconducting phase. For $\phi=\pi/2$, the Fourier transform of the pairing Hamiltonian $H_{\rm pair}$ produces momentum-dependent pairing potentials proportional to $\sin(k_x)$ for spin-$\uparrow$ and $\sin( k_{y})$ for spin-$\downarrow$ fermions. This leads to the $p$-wave structure as in Eq.\ \ref{pair_p} near the Fermi surface.

%{\bf Comment: I am not sure if $H_{\rm pair}$ should have a factor of $i\Delta_0$ instead of $\Delta_0$. How do you get a odd-momentum out of your pairing term? Needs a bit of explanation. }
%
%\textcolor{green}{\underline{Comment}:This pairing hamiltonian correspond to pairing potential of form like $\Delta_{\uparrow \uparrow}\propto \sin(k_x)$. But the pairing potential given in Eq.\ref{pair_p} is structurally different. Any one needs to be changed.}
%

We consider now the full Hamiltonian $H + H_{\text{pair}}$ from Eqs.\ \ref{ham1} and \ref{hpair} with open-boundary conditions along $x$-direction, keeping translation invariance along $y$ intact. As the carriers along the $x$-direction are spin-$\uparrow$ fermions, we look at the spin-$\uparrow$ sector of the full Hamiltonian. On Fourier transforming along $y$-direction, the Hamiltonian decouples into independent $ k_{y}$ sectors given by
\begin{eqnarray}
    H_{\uparrow} &=& \sum_{ k_{y}} H( k_{y}) = \sum_{ k_{y}} \sum_{j_x} \left[ \varepsilon( k_{y})\,c^\dag_{\uparrow}(j_x,k_{y})c_{\uparrow}(j_x, k_{y}) \right. \nonumber \\
    && - (t + t_R) \big(c^\dag_{\uparrow}(j_x + 1, k_{y})c_{\uparrow}(j_x, k_{y})+\mathrm{h.c.} \big) \nonumber \\
    && \left. + \big(\Delta_0 e^{i \phi} \, c^\dag_{\uparrow}(j_x+1, k_{y})c^\dag_{\uparrow}(j_x, k_{y})+\mathrm{h.c.}\big) \right],
    \label{hkitaev}
\end{eqnarray}
where 
\begin{equation}
\varepsilon( k_{y}) = -2\,(t-t_R)\cos k_{y} - \mu.
\end{equation}
is the the momentum-dependent onsite potential. Notice that for each conserved momentum $ k_{y}$, $H(k_{y})$ is equivalent to a 1D $p$-wave superconductor (a Kitaev wire)~\cite{kit1} with nearest-neighbor hopping $\tilde t = t + t_R$, $p$-wave pairing amplitude $\Delta_0$ and effective chemical potential $\varepsilon( k_{y})$. Upon Fourier transforming along $x$ and introducing two-component Nambu operators $\Psi^\dag(k_x) \equiv (c^\dag_{\uparrow}(k_x, k_{y}), \, c_{\uparrow}(-k_x, k_{y}))$, we can write $H( k_{y})$ in the standard Boguliubov-de Gennes form:
\begin{equation}
    H( k_{y}) = \frac{1}{2} \sum_{k_x} \Psi^\dag(k_x) \mathcal{H}(k_x) \Psi(k_x), 
\end{equation}
where 
\begin{equation}
    \mathcal{H}(k_x) = \begin{pmatrix}
        \xi(k_x,  k_{y}) & \tilde{\Delta}(k_x) \\
        \tilde{\Delta}^*(k_x) & -\xi(k_x,  k_{y}) \\
    \end{pmatrix}
\end{equation}
with $\xi(k_x,  k_{y}) = \varepsilon( k_{y}) - 2(t + t_R) \cos(k_x)$ the kinetic energy and $\tilde{\Delta}(k_x) = -2i\Delta_0 e^{i\phi} \sin(k_x)$ the momentum-space pairing potential. We note that, for $\phi=\pi/2$, the form matches with the analytical form Eq.~(\ref{pair_p}) for small values of momentum. From the criterion for a Kitaev wire being in the topological phase, the condition for which each $ k_{y}$-dependent slice is a topological superconductor is given by
\begin{equation}
    2 (t + t_R) > |2\,(t-t_R)\cos  k_{y} + \mu|, \nonumber
\end{equation}
which always holds for $\mu = 0$ and $t_R / t > 0$. Thus each effective 1D superconductor hosts zero-energy Majorana edge modes, which we briefly discuss in the Appendix \ref{app:majorana}.

\subsection {Josephson response}
\label{sec:jr} 

%------------------------------
\begin{figure}
 	\begin{center}
    \includegraphics[width=\linewidth]{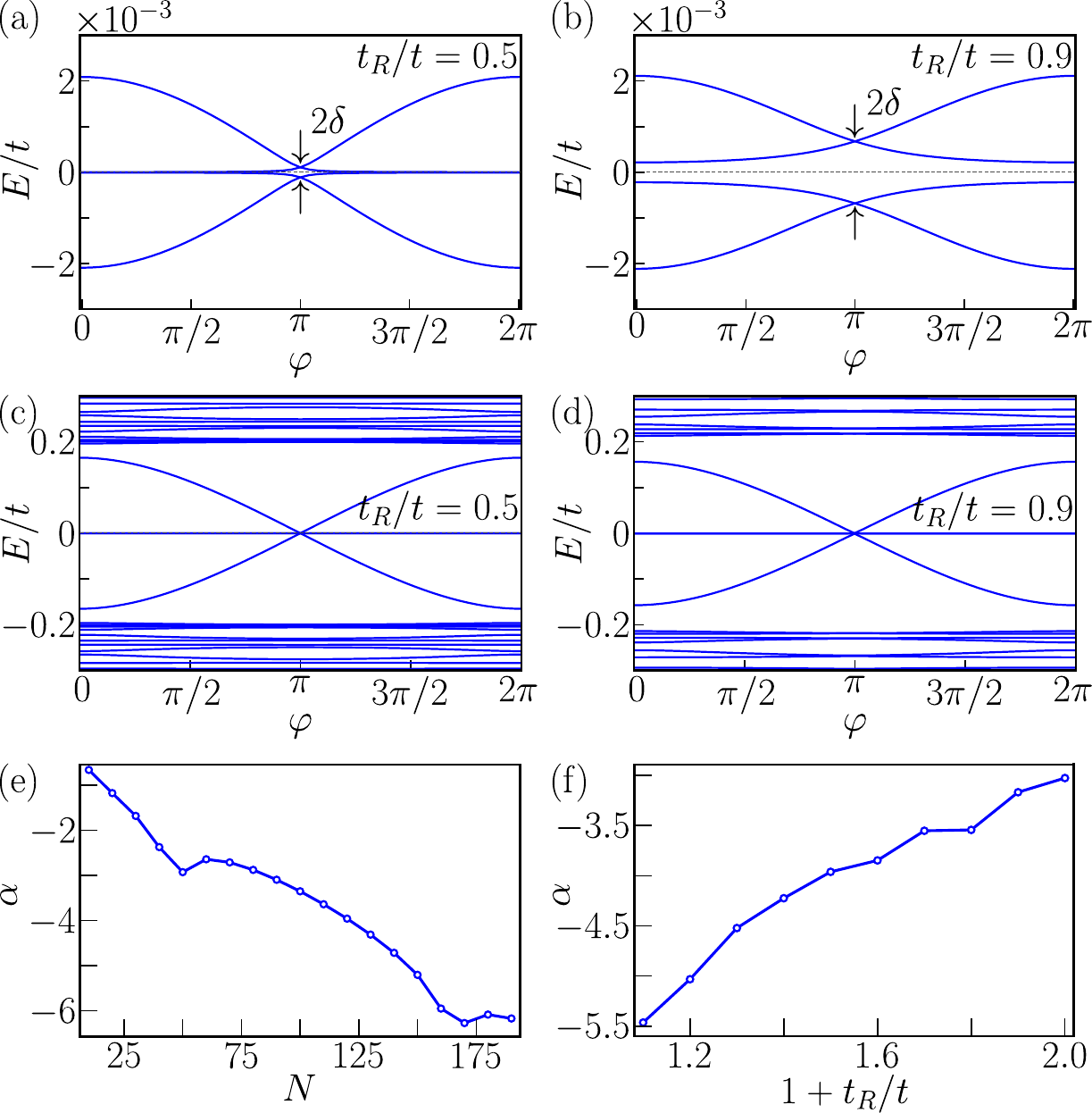}
 \caption{(a) - (d) Andreev bound state spectrum of the Josephson junction between two effective 1D superconductors with fixed $ k_{y} = 0$ and weak-link hopping $t_J$. The number of sites of each superconductor is $N=120$ along the $x$-direction, $\Delta_0 / t = 0.1$ and $\mu / t = 0$. We consider two limiting cases - (a), (b) AB limit for which $t_J/t = 0.01$ and (c), (d) KO limit for which $t_J / t = 1.0$. The corresponding $t_R / t$ values are shown in the figures. The subgap energy spectrum versus $\varphi$ shows a gap of $2\delta$ at $\varphi = \pi$, where $\delta$ in the energy splitting between Majorana edge modes in each superconductor. This gap increases with $t_R$ in the AB limit but is almost independent of it in the KO limit for $N=120$.  (e) Plot of  $\alpha \equiv \mathrm{log}_{10}(\delta / t)$ as a function of $N$, confirming exponential decay of $\delta$ with $N$. (f) Plot of $\alpha$ with $1 + t_R / t$ showing linear scaling of coherence length $\xi \sim \alpha$ with $t_R$. For both plots (e) and (f) $t_J/t=0.01$. 
 \label{abs} }
 	\end{center}	
 \end{figure}

In this section, we consider a short, transparent junction connecting two such superconducting leads with phase difference $\varphi$. For each transverse momentum $ k_{y}$, the analysis of Sec.~\ref{ncs1} applies. Witten’s index theorem then guarantees bound states with energies $E\propto \pm \cos(\varphi/2)$, with a protected crossing at $\varphi=\pi$ for each $ k_{y}$. Importantly, one expects that a weak scalar interfacial scattering shall not hybridize them and the $4\pi$ nature will persist at small barrier strengths.  When parity is not explicitly conserved, the Josephson current becomes 2$\pi$ periodic, but with a sharp jump at $\varphi=\pi$, indicating parity switch. Moreover, in this regime and in a junction with strong barrier potential (near the AB limit) the critical current depends linearly on the tunneling amplitude $t_J$ between the two superconductors for a $p$-wave junction with $4\pi$ periodic response; this distinguishes such junctions from the standard $s$-wave $2\pi$ periodic case where the critical current has quadratic dependence on the tunneling amplitude \cite{jiang2011unconventional,kumari2024josephson}.  

In what follows, we shall validate these expectations numerically. First, we note that with increasing $t_R/t$ (for fixed, small $\Delta_0$), the localization length of the edge-modes for a given superconductor  grows linearly with $t_R$: $\xi\propto (t+t_R)/\Delta_0$ ~\cite{kit1}. %Following ~\cite{kit1}, we find the localisation length for the Majorana's in each superconductor to be $\xi = 2 / \mathrm{log}(\frac{t + t_R - \Delta_0}{t + t_R + \Delta_0}). For $\Delta_0 \ll t_R$, we can expand the logarithm to get $\xi \simeq (t + t_R) / \Delta_0$. The splitting between the Majoranas goes as $\delta \propto e^{-N / \xi}$, where $N$ is the system size.%
This results in larger overlap of the wave-functions of opposite-edge Majoranas in each lead, increasing their finite-size splitting $2\delta$. Moreover, in a Josephson junction between two topological superconductors with a weak-link hopping $t_J$, the edge-modes at the junction hybridizes, giving rise to Andreev spectrum. At the superconducting phase difference of $\varphi=\pi$, the bound states at the junction and at two extreme edges exchange due to parity reversal. This crossing is protected by the fact that parity of each superconductor remains a good quantum number, and is responsible for the 4$\pi$ nature of the resulting current-phase relationship (CPR), as long as $\delta \ll t_J$.

%--------------------------
\begin{figure} [!htbp]
 	\begin{center}
 		\includegraphics[width=1.0\linewidth]{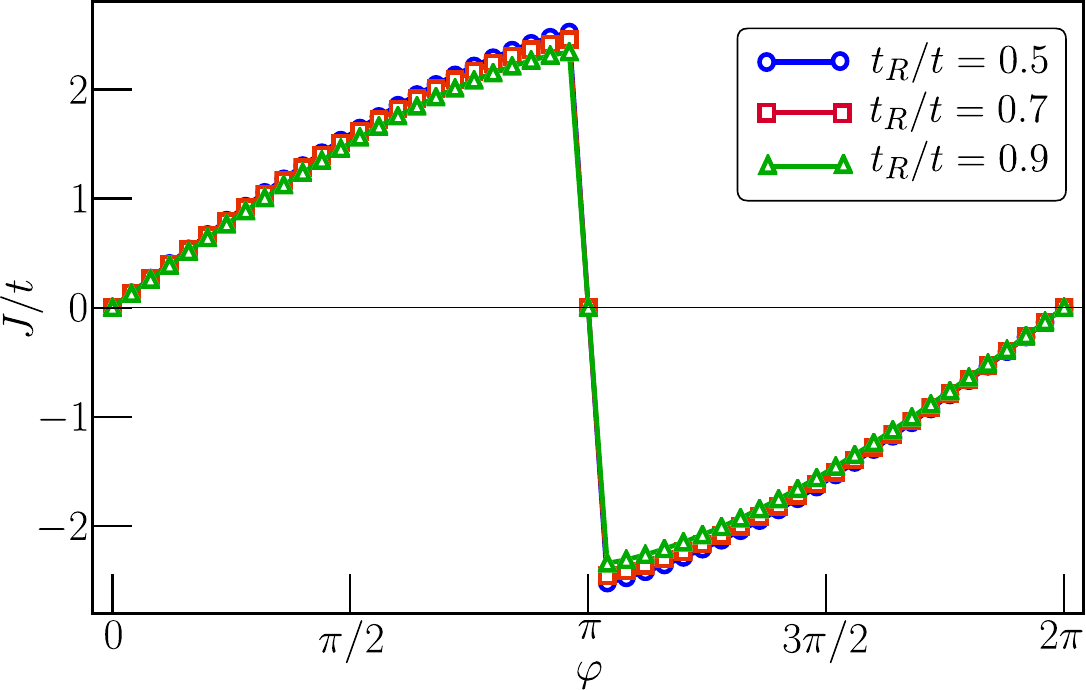}
 \caption{The current-phase relationship for the spin-$\uparrow$ current in the Josephson junction between two 2D altermagnet superconductors for junction hopping $t_J / t = 1.0$, shown for a few values of $t_R / t$. The number of sites of each superconductor is $N=120$ along $x$-direction, $W = 10$ along $y$-direction. For all plots, $\Delta_0 / t = 0.1$ and $\mu / t = 0$. The Josephson current remains $4\pi$-periodic with a sharp jump at $\varphi=\pi$ for all values of $t_R / t$. \label{josphi2} }
 	\end{center}	
 \end{figure}
%-----------------------------

We depict this numerically in Fig.\ \ref{abs}, where we construct a Josephson junction of two such superconductors, each of $N$ sites along $x$-direction, keeping translation invariance along $y$ direction. We plot the typical sub-gap spectrum for a given transverse momentum ($ k_{y}$), where we observe that with increasing $t_R/t$, $\delta$ also increases exponentially. The sub-gap spectrum has been shown for both the limiting cases of the junction hopping, $t_J/ t = 0.01$ (strong barrier or the AB limit) and $t_J/t = 1.0$ (weak barrier or the KO limit). We also recover the state switching at $\varphi=\pi$. 

We now consider a full 2D system of two superconductors, each of finite width $W$ along the $y$-direction and length $N$ along $x$-direction, with a weak-link hopping $t_J$ at the junction and calculate the Josephson current $J(\varphi)$ using the non-equilibrium Green-function method of Appendix.~\ref{gfmeth}. Numerically $J(\varphi)$ is computed as a sum of the bond currents that flows from the left to the right superconductor across the weak link. As the junction is along the $x$-direction, the Fermi lines parallel to the $ k_{y}$-axis hosts only spin-$\uparrow$ fermions and the Josephson current is carried predominantly by them (the spin-$\downarrow$ states are nearly insulating along $x$ when $t_R/t$ is large). 
%--------------------------
\begin{figure} [!htbp]
 	\begin{center}
 		\includegraphics[width=1.0\linewidth]{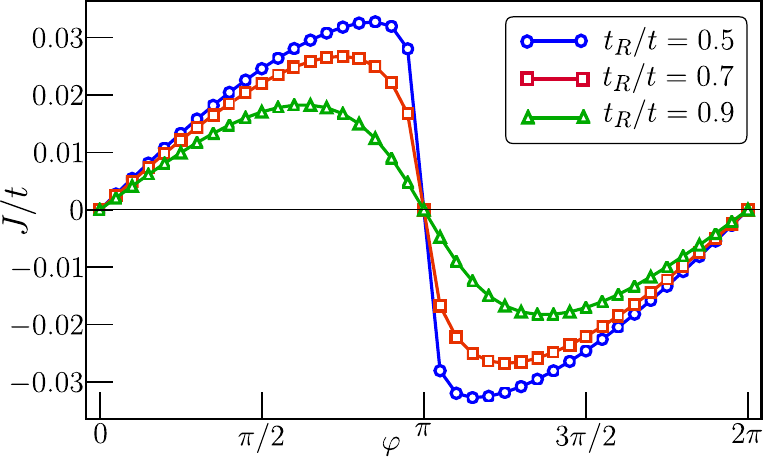}
 \caption{Current-phase relationship for the spin-$\uparrow$ current in the Josephson junction between two 2D altermagnet superconductors for junction hopping $t_J / t = 0.01$, shown for a few values of $t_R / t$. The number of sites of each superconductor is $N=120$ along $x$-direction, $W = 10$ along $y$-direction.  For all plots,  $\Delta_0 / t = 0.1$ and $\mu / t = 0$. With decrease in $t_R / t$, the Josephson current $J$ (in units of $t$) changes from almost $2\pi$-periodic to $4\pi$-periodic, along with a discontinuity at $\varphi=\pi$ where the fermion parity switches sign. }\label{josphi} 
 	\end{center}	
 \end{figure}
%-----------------------------
%----------------------------- 
The CPRs for the $t_J / t = 1$ case (near the KO limit) are shown in Fig.\ \ref{josphi2}, showing sharp $4\pi$ periodicity proportional to $\sin(\varphi / 2)$ for all values of $t_R/t$, with the characteristic jump at $\varphi=0$ due to the switching of the fermion parity. Increasing $t_R/t$ leads to small decrease in the current as the localization length of the edge modes of each superconductor increases. For the small hopping limit $t_J / t = 0.01$ (AB limit), resulting CPRs are shown in Fig.\ \ref{josphi}, where we note that, with increasing $t_R/t$, the CPR goes from a nearly $4\pi$ periodic form proportional to $\sin(\varphi / 2)$ (noticed especially for $t_R/t=0.5$), to a $2\pi$ periodic form proportional to $\sin(\varphi)$. We note that, as with increase in $t_R/t$, the localization length of the edge modes of individual superconductors also increases exponentially as discussed earlier. Thus to see a sharper jump at $\varphi=\pi$, one has to go to larger system size. These results indicate that the $4\pi$ periodicity holds both in the AB ($t_J/t \ll 1$) and the KO ($t_J/t \simeq 1$) limits albeit with strong finite size corrections in the former case. These results are therefore consistent with the predicted barrier potential independence as discussed in Sec.\ \ref{ssec2}.   

\begin{figure} [!htbp]
 	\begin{center}
 		\includegraphics[width=1.0\linewidth]{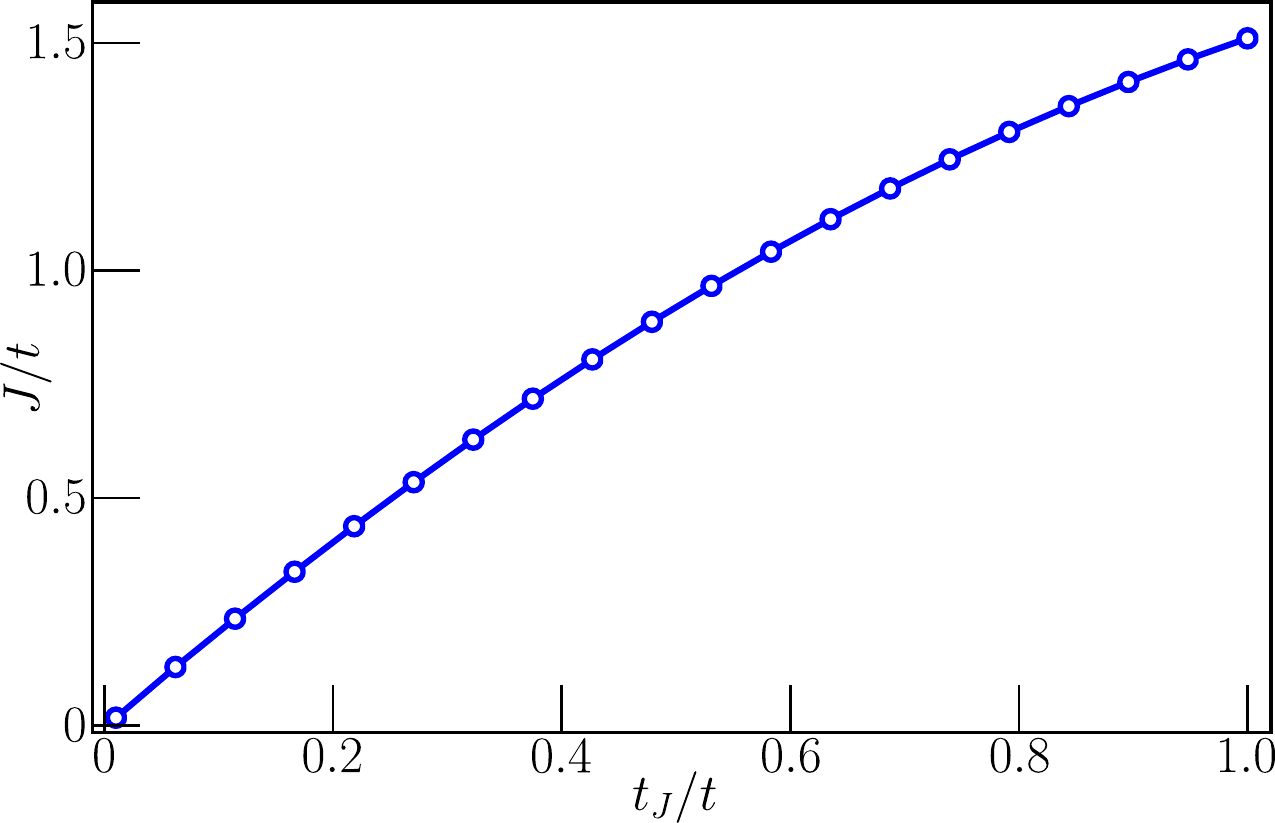}
 \caption{Plot of Josephson junction spin-$\uparrow$ current $J$ (in units of $t$) for a junction between two 2D superconductors as a function of $t_J / t$ for fixed phase difference $\varphi = \pi / 2$. The number of sites of each superconductor is $N=120$ along $x$-direction and $W = 10$ along $y$-direction.  For this plot, we have used $\Delta_0 / t = 0.1$, $\mu / t = 0$ and $t_R / t = 0.9$. The current scales linearly with $t_J$ for small $t_J \le \Delta_0$. } \label{josjunc} 
 	\end{center}	
 \end{figure}

When the periodicity of Josephson current is close to a $4\pi$, the variation of the spin-$\uparrow$ Josephson current versus the junction coupling $t_J$ at some fixed $\varphi = \pi / 2$ is shown in Fig.\ \ref{josjunc}. We find a linear dependence of the Josephson current on $t_J$ for small $t_J \le \Delta_0$ \cite{kumari2024josephson}; this situation is therefore qualitatively different from the expected quadratic dependence of $J$ on $t_J$ for $2\pi$ periodic junctions. For larger $t_J/t$ the current is expected to saturate and this behavior is seen in Fig.\ \ref{josjunc}. These results therefore points towards robust $4\pi$ nature of the Josephson current even in the AB limit. 

%The inset in Fig.\ \ref{josjunc} shows the exponent $\chi = d \log J / d \log t_J$ against $t_J$, which shows all the points being close to unity, consistent with the expectation from the analysis of Sec.\ \ref{ncs1} in the KO limit.}%

\begin{figure}[!htbp]
 	\begin{center}
 		\includegraphics[width=1.0\linewidth]{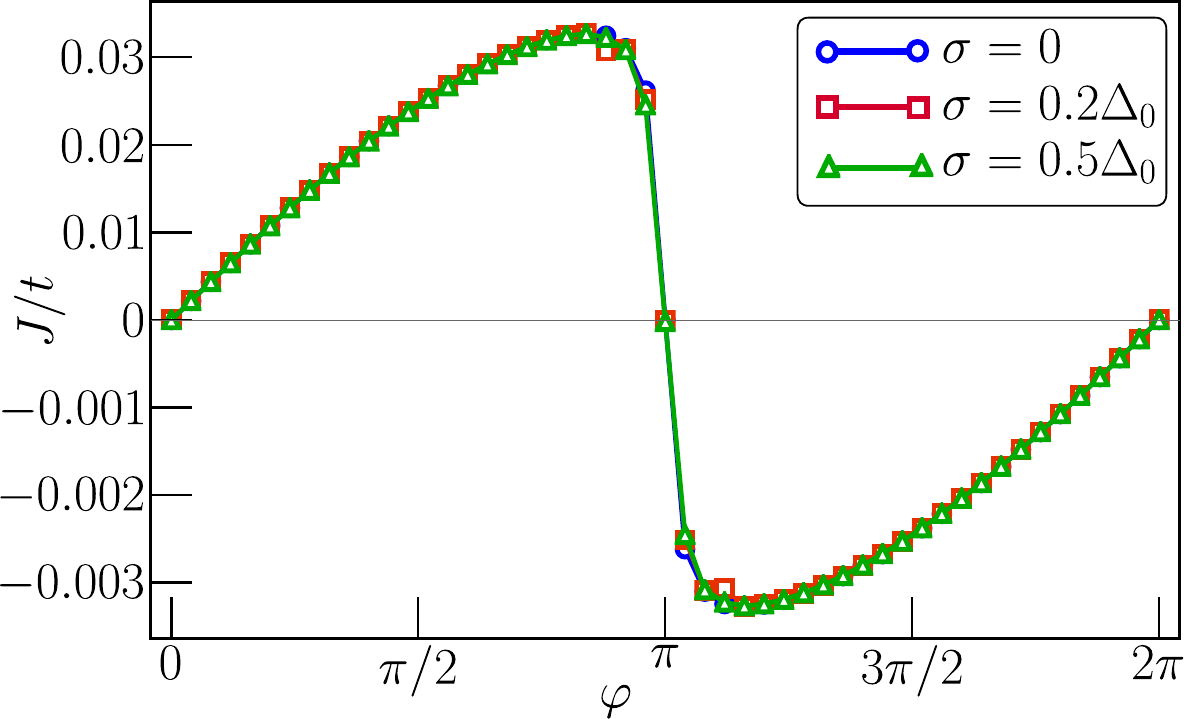}
 \caption{The disorder averaged spin-$\uparrow$ current versus phase difference $\varphi$, in the Josephson junction between two 2D superconductors with edge disorder, for few values of the disorder strength $\sigma$. The number of sites of each superconductor $N=120$ along $x$-direction and $W = 10$ along $y$-direction. For all plots,  $\Delta_0 / t = 0.1$, $\mu / t = 0$, $t_R / t = 0.5$ and $t_J / t = 0.01$. For each disorder strength, we average over three disorder configurations. Note that for weak disorder $\sigma / \Delta_0 \ll 1$, the $4\pi$-periodic nature of the CPR as well as the discontinuity at $\varphi = \pi$ remains intact.\label{josdis} }
 	\end{center}	
 \end{figure}

To examine the robustness of the boundary modes and thus the 4$\pi$ nature of the junction against impurities at the junction, we add localized disorder potential to the sites along the two adjacent edges of the superconductors. The potential along these sites are chosen from a Gaussian distribution with zero mean and standard deviation $\sigma$, which characterizes the strength of the potential. The configuration averaged Josephson current versus $\varphi$ plots are shown in Fig.\ \ref{josdis} for fixed $t_J/t=0.01$ and for several $t_R/t$ and $\sigma$. For weak disorder, $\sigma/\Delta_0 < 1$, the $4\pi$ character and the jump near $\varphi=\pi$ are essentially unchanged; this is a consequence of the fact that in this regime the disorder fails to hybridize the subgap states. This feature is consistent with the discussion in Sec.~\ref{ssec2} where such a lack of hybridization was predicted due to topological protection of the $p$-wave subgap Andreev bound states against impurity induced scattering. For larger $\sigma$, it is expected that the modes eventually mix, gradually restoring a $2\pi$-periodic CPR; however, for disorder localized at the edges this requires a very large $\sigma$ since such a potential can not effectively couple between localized Andreev modes with delocalized extended states above the gap.  This robustness, although not shown here, is independent of the $t_J/t$ ratio.

%-------------------------------
%-------------------------------
%-------------------------------

\section{Discussion}
\label{sec:diss}
In this work, we have analyzed the Andreev bound states of tunnel junctions using supersymmetric quantum mechanics, Our main finding is that these bound states, in the transparent junction limit, can be viewed as consequence of the index theorem. Our analysis shows that the BdG equations governing the superconducting quasiparticle in such junctions can be cast into a supersymmetric form with a part of the pair-potential acting as the superpotential. We have shown that this superpotential necessarily changes sign as one traverses through the junction; the resultant bound states, whose existence is guaranteed by the index theorem, are precisely the Andreev states that occur in a Josephson junction. Our analysis holds for both non-chiral and chiral superconductors. It shows that these bound states are robust against spatial structure of the pair-potential; their existence depend only on its asymptotic properties. We note that the BdG equations of the transparent Josephson junctions serve as a rare example supersymmetric quantum mechanical systems where the bound states originating from index theorem can occur at finite energy. 

Our analysis provides a natural route to perturbative study of the effect of a barrier potential or charged non-magnetic impurities in such junctions. We show that the Andreev bound states of non-chiral $p$-wave junctions exhibit a topological robustness against hybridization due to such potentials or impurities; this result directly follows from the structure of their wavefunction and do not depend on the precise form of the barrier or impurity potential. This robustness is key to $4\pi$ periodic Josephson effect of such junctions; it is a feature that is not shared by either $s$-wave or chiral superconductors whose bound states have a different structure allowing for hybridization due to barrier or impurity potential and consequently leading to the more common $2\pi$ periodic Josephson current. Our analysis shows that this topological robustness is more general than the context of 1D Kitaev chain where it is discussed as a consequence of Majorana bound states at the end of the chain (App.\ \ref{app:majorana}).

We have tested this analytic understanding using a microscopic tight-binding model describing a 2D altermagnet with equal-spin triplet $p$-wave pairing. For this model we used a non-equilibrium Green function formalism to compute spin-resolved transport in a Josephson geometry (Appendix.~\ref{gfmeth}), where we have studied several features of the current phase relationship both in the KO and the AB limit. Our results seem to be consistent with the presence of robust $p$-wave–like Andreev bound states in both these limits; this agrees with the prediction of the independence of the $4\pi$ periodic Josephson current on the barrier potential. We further analyzed the dependence of the Josephson current on the junction hopping $t_J$ and showed that, in the short-junction limit and for $t_J/t \le \Delta_0$, it exhibits a linear scaling $t_J/t$ consistent with $4\pi$ periodic Josephson response. The inclusion of edge disorder confirmed that the qualitative features of the spin-polarized Josephson response and the effective $4\pi$ periodicity remain robust up to moderate disorder strengths, in line with the index-theorem argument.

The present analysis suggests several directions for future work. Our exact solution of the Andreev bound state in the transparent junction limit is expected to provide a natural starting point for several perturbative studies such as   analysis of Josephson junctions with weak magnetic impurities at the interface. This method can also possibly be generalized to include junctions between superconductors with different pairing symmetries. The applicability of this method to longer weak links and to multi-terminal Josephson junctions also deserves further study. Our numerical results predicts spin polatrized Josephson current in superconducting altermagnets. Moreover our study shows that Josephson junctions of two such altermagnets should display $4 \pi$ periodic Josephson current which is robust against weak junction disorder; this effect should manifest itself in absence of odd Shapiro steps \cite{shap1,shap2,shap3} and can be experimentally tested. Further, we predict that the spin polarization of the Josephson current should be sensitive to the orientation of the junction for such altermagnets which may also be experimentally verified. 

In conclusion we have provided an index-theoretic route to analyzing Andreev bound states in Josephson junctions. Our analysis brings out the topological nature of the robutness of the $4\pi$ Josephson effect for $p$-wave junctions. We have numerically rested our results using superconducting altermagnets as a possible platform and have pointed out experimental signature that may corroborate our theoretical results.  

\section{Acknowledgment} 

KS thanks DST, India for support through SERB project JCB/2021/000030.

\color{black}
\appendix
\section{Majorana description of the junction bound states}
\label{app:majorana}

For completeness, we briefly recast the short-junction problem in terms of Majorana modes and relate this to the BdG spinor structure used in Sec.~\ref{ncs1}. For clarity we focus on a single topological $p$-wave channel along the junction direction; the spin and transverse-momentum labels of the main text can be restored straightforwardly.

We consider two semi-infinite topological $p$-wave superconductors forming a short Josephson junction with phase difference $\varphi$. In the topological regime, each lead contributes a Majorana zero mode localized at the contact, which we denote by $\gamma_L$ and $\gamma_R$ for the left and right superconductors, respectively. The low energy effective Hamiltonian describes the overlap of these modes across a short junction~\cite{jiang2011unconventional}
\begin{equation}
	H_{\rm eff}(\varphi)= i\,\varepsilon(\varphi)\,\gamma_L\gamma_R,\qquad 
	\varepsilon(\varphi)=E_M\cos\!\frac{\varphi}{2}.
	\label{eq:Heff_app}
\end{equation}
Equation~\eqref{eq:Heff_app} is the Majorana representation of the subgap Andreev problem whose BdG form was analysed in Sec.~\ref{ncs1}. It is convenient to combine the two Majoranas into a single complex fermion
\begin{equation}
	f=\tfrac12(\gamma_L+i\gamma_R),\qquad f^{\dagger}=\tfrac12(\gamma_L-i\gamma_R).
	\label{eq:f_app}
\end{equation}
Using $\{\gamma_\mu,\gamma_\nu\}=2\delta_{\mu\nu}$ and  $i\gamma_L\gamma_R=2f^\dagger f-1$,
\begin{equation}
	H_{\rm eff}(\varphi)
	=\varepsilon(\varphi)\,\big(2f^\dagger f-1\big).
		\label{eq:energies_app}
\end{equation}
This Hamiltonian has eigenvalues $E(\varphi) = \varepsilon(\varphi)$ and the corresponding eigenstates are $|-\rangle=f|\Omega\rangle$ and $|+\rangle=f^{\dagger}|\Omega\rangle$, where $|\Omega\rangle$ is the state annihilated by $f$.  For fixed fermion parity $n=\langle f^\dagger f\rangle=0,1$, the Josephson current is written as
\begin{equation}
	I_n(\varphi)=\frac{2e}{\hbar}\,\partial_\varphi\!\big[(2n-1)\varepsilon(\varphi)\big]
	=\mp\,\frac{eE_M}{\hbar}\,\sin\!\frac{\varphi}{2},
\end{equation}
where the upper (lower) sign corresponds to $n=1$ ($n=0$). Because $\sin(\varphi/2)$ is $4\pi$-periodic, an adiabatic evolution within a fixed-parity sector produces a $4\pi$-periodic current–phase relation, in agreement with the BdG analysis of the $p$-wave junction in Sec.~\ref{ncs1}. If fermion parity relaxes on the Josephson time scale, the system instead follows the instantaneous ground state $E(\varphi)=-|\varepsilon(\varphi)|$, and the current becomes $2\pi$-periodic with a cusp at $\varphi=\pi$.

To connect this Majorana description with the Nambu spinors used in the main text, we now embed the Majorana operators into the local BdG eigenmode at the junction. Let $\psi_J$ denote the fermionic quasiparticle operator for the localized Andreev bound state obtained from the BdG equation in Sec.~\ref{ncs1}. Up to a real, positive envelope near the contact, the Majoranas at the left and right ends can be written as electron–hole superpositions with the condensate phases $\theta_L$ and $\theta_R$ of the two leads,
\begin{align}
	&\gamma_L \simeq e^{-i\theta_L/2}\psi_J+e^{+i\theta_L/2}\psi_J^\dagger, \nonumber \\
	&\gamma_R \simeq i\!\left(e^{-i\theta_R/2}\psi_J - e^{+i\theta_R/2}\psi_J^\dagger\right).\nonumber
\end{align}
Inserting this into Eq.~\eqref{eq:f_app} and expanding in the BdG basis, one finds for the positive-energy Andreev mode
\begin{equation}
	f \propto \tfrac12(\gamma_L+i\gamma_R)
	= u_1\psi_J + v_1\psi_J^\dagger,
\end{equation}
with
\begin{equation}
	\frac{u_1}{v_1} =  \frac{e^{-i\theta_L} + i e^{-i\theta_R}}{e^{i\theta_L} + i e^{i\theta_R}} \equiv e^{i\chi}.
\end{equation}
Thus, in the BdG basis, the state $f|\Omega\rangle$ is written a spinor, as \[d_{-} = \left(\begin{array}{c}
	u_1 \\
	v_1
\end{array}\right) \propto \left(\begin{array}{c}
e^{i\chi} \\
1
\end{array}\right).\]
Similarly, for the other state,
	\begin{equation}
	f^{\dagger}\propto \tfrac12(\gamma_L-i\gamma_R) = u_2\psi_J+v_2\psi_J^\dagger.\nonumber
\end{equation}
This gives the ratio \begin{align*}
	\frac{u_2}{v_2} &= \frac{e^{-i\theta_L} - i e^{-i\theta_R}}{e^{i\theta_L}  -i e^{i\theta_R}} \nonumber\\
    &= \left(\frac{e^{i\theta_L} + i e^{i\theta_R}}{e^{-i\theta_L}  +i e^{-i\theta_R}}\right)^{*} = \left(e^{-i\chi}\right)^* = e^{i\chi}.
\end{align*}
Here, the fact that $\frac{u_1}{u_2} = \frac{v_2^*}{u_2^*}$ is simply a consequence of particle-hole symmetry, where the operator is $\Sigma = \sigma_x C$. Thus, in the BdG basis, the state $f^{\dagger}|\Omega\rangle$ is written a spinor, as \[d_{+} = \left(\begin{array}{c}
	u_2 \\
	v_2
\end{array}\right) \propto \left(\begin{array}{c}e^{i\chi} \\
1
\end{array}\right).\]
Importantly, the two Andreev branches related by $E\to -E$ share the same local Nambu structure (differing only by a global phase), so that a weak scalar impurity at the junction, coupling to $\psi_J^\dagger\psi_J\propto\tau_3$ in Nambu space, has vanishing matrix element between them.

\section{Details of transport simulation}
\label{gfmeth} 
 
For a superconducting system coupled to superconducting/non-superconducting reservoirs, we employ the Bogoliubov-de Gennes (BdG) formalism with the Nambu spinor basis $\Psi = (c, c^{\dagger})^T$, where $c$($c^\dagger$) is the electron annihilation(creation) operator of the system. The full Hamiltonian can be written as $H = H_S + H_L + H_C$, with Hamiltonian $H_S$ for the central superconducting region given by 
\begin{eqnarray}
 	H_{\rm S} &=& \frac{1}{2} \sum_{lm} \begin{pmatrix} c_l^{\dagger} & c_l \end{pmatrix}
 	\begin{pmatrix}
 		A_{lm} & \Delta_{lm} \\
 		\Delta_{lm}^* & -A_{lm}^*
 	\end{pmatrix}
 	\begin{pmatrix} c_m \\ c_m^{\dagger} \end{pmatrix} \nonumber\\ 
 	&\equiv& \frac{1}{2} \Psi^{\dagger} \check{A} \Psi, \label{numformeq1} 
 \end{eqnarray}
where $\check{A} = \begin{pmatrix} A & \Delta \\ \Delta^{\dagger} & -A^T \end{pmatrix}$ is the BdG Hamiltonian matrix. For the lead $\lambda= L, R$, the Hamiltonian $H_{L}^{\lambda}$ is given by 
 \begin{eqnarray}
 	H_{\rm L}^{\lambda} &=& \frac{1}{2} \sum_{\alpha\beta} \begin{pmatrix} a_{\alpha}^{\lambda\dagger} & a_{\alpha}^{\lambda} \end{pmatrix}
 	\begin{pmatrix}
 		F_{\alpha\beta}^{\lambda} & \Delta_{\alpha\beta}^{\lambda} \\
 		\Delta_{\alpha\beta}^{\lambda} & -F_{\alpha\beta}^{\lambda}
 	\end{pmatrix}
 	\begin{pmatrix} a_{\beta}^{\lambda} \\ a_{\beta}^{\lambda\dagger} \end{pmatrix}\nonumber\\
 	&\equiv& \frac{1}{2} \phi_{\lambda}^{\dagger} \check{F}^{\lambda} \phi_{\lambda}\label{numformeq2} 
 \end{eqnarray}
 where $\phi_{\lambda} = (a^\lambda, {a^\lambda}^{\dagger})^T$ and $a^\lambda$(${a^\lambda}^{\dagger}$) is the electron annihilation(creation) operator of the lead $\lambda$.
 Finally, the coupling Hamiltonian $H_c^{\lambda}$ between the lead $\lambda$ and the central superconductor is given by 
 \begin{eqnarray}
 	H_{\rm C}^{\lambda} &=& \sum_{\alpha l} \left( K_{\alpha l}^{\lambda} a_{\alpha}^{\lambda\dagger} c_{l} + \text{h.c.} \right)
 	\equiv a^{\lambda\dagger} K^{\lambda} c + \text{h.c.}, \label{numformeq3} 
 \end{eqnarray}
which in the Nambu space can be written as 
 \begin{eqnarray}
 	H_{\rm C}^{\lambda} &=& \frac{1}{2} \phi_{\lambda}^{\dagger} \check{K}^{\lambda} \Psi + \text{h.c.}, \,\,
 \check{K}^{\lambda} = \begin{pmatrix} K^{\lambda} & 0 \\ 0 & -(K^{\lambda})^* \end{pmatrix}. \label{numformeq4} 
 \end{eqnarray}
 The equations of motion for the Nambu spinors for both the leads and the superconducting system, are given respectively by
 \begin{eqnarray}
 i \dot{\phi}^\lambda(t) &=&
 	\check{F}^\lambda \phi^\lambda(t) + \check{K}^\lambda{}^\dagger \Psi(t)\nonumber\\
 i \dot{\Psi}(t) &=& \check{A} \Psi(t) 
 + \sum_\lambda \check{K}^\lambda \phi^\lambda(t). \label{numformeq5} 
 \end{eqnarray}
 The first of the two equations in Eq.\ \ref{numformeq5} can be formally solved to yield 
 \begin{eqnarray}
 	\phi^\lambda(t) &=& i \bg(t - t_0) \phi^\lambda(t_0) + \int_{t_0}^t dt'\, \bg(t-t') \bK{}^\dagger \bpsi(t'), \nonumber\\ \label{numformeq6} 
 \end{eqnarray}
 where $t_0$ is assumed to be the distant past when the coupling is switched on. Here $g_{\lambda}(t-t')$ is the Fourier transform of the retarded Green's function $g_{\lambda}^r$ of the leads which is given by 
 \begin{eqnarray}
 	\check{g}_{\lambda}^r(\omega) &=& \left[ \omega \check{I} - \check{F}^{\lambda} \right]^{-1}. \label{numformeq7} 
 \end{eqnarray}

In terms of these Green's function, one can write 
 \begin{eqnarray}
&& [i\partial_t - \bA]\bpsi(t) + i \int_{t_0}^t dt'\, \check{\Gamma}(t-t')\bpsi(t') = \sum_\lambda \bK \xi_\lambda(t), \nonumber\\
 && \bxi(t) =  i \bg(t-t_0) \bphi(t_0), \nonumber\\
 && \bGamma(t-t') = i \sum_\lambda \bK \bg(t-t') \bK{}^\dagger.\label{numformeq8} 
 \end{eqnarray}  
We note that $\Gamma$ denotes the self-energy matrix. In terms of these, the equation for the Green's function equation can be written in Fourier space as
 $(\omega - \bA + i \bGamma(\omega))\bG (\omega) = \check{I}$. This finally leads to 
 \begin{eqnarray}
 	\bpsi(t) &=& \sum_\lambda \int \frac{d\omega}{2\pi} \, e^{-i\omega t} \bG(\omega) \bK \bxi(\omega). \label{numformeq9} 
 \end{eqnarray}
We numerically solve Eq.\ \ref{numformeq7} and use it to obtain the 
 \begin{eqnarray} 
 && \langle \xi^{\lambda\dagger}_{y\eta}(\omega)\,
 	\xi^{\lambda'}_{y'\eta'}(\omega') \rangle = \mathrm{Tr}\!\left[\rho\,
 	\xi^{\lambda\dagger}_{y\eta}(\omega)\,
 	\xi^{\lambda'}_{y'\eta'}(\omega')\right] \nonumber \\
&& = (2\pi)^2 \delta_{\lambda\lambda'} \delta_{\eta\eta'} 
 	\delta(\omega-\omega')\, \rho^\lambda_{y'y}\,
 	f^\lambda(\omega-\mu^\lambda). \label{numformeq10} 
 \end{eqnarray}
 
In what follows, we shall be interested in computing the current flowing through the junction~\cite{mukherjee2017transport,kumari2024josephson}. To this end, we define the bond current operator. This is most easily done by defining a generalized hopping term in the Hamiltonian given by 
 \begin{eqnarray}
 	H^{j \to i}(\theta) &=& \frac{1}{2} \Psi_i^{\dagger} \check{t}_{ij} e^{-i\theta} \Psi_j + \text{h.c.} \label{numformeq11}
 \end{eqnarray}
 where $\check{t}_{ij} = \begin{pmatrix} t_{ij} & 0 \\ 0 & -(t_{ij})^* \end{pmatrix}$ is the hopping matrix in Nambu space. The current operator is obtained by differentiating $H$ with respect to the phase $\varphi$ and is given by 
 \begin{eqnarray}
 	j_{j \to i} &=&  \left. \frac{\partial H^{j \to i}(\theta)}{\partial \theta} \right|_{\theta=0}
 	= -\frac{i}{2} \left( \Psi_i^{\dagger} \check{t}_{ij} \Psi_j - \Psi_j^{\dagger} \check{t}_{ij}^{\dagger} \Psi_i \right). \nonumber\\ \label{numformeq12}
 \end{eqnarray}

We note here that $j$ given in Eq.\ \ref{numformeq12} accounts for both electron-like and hole-like contributions.	
The expectation value of the current operator given by 
 \begin{eqnarray}
 	\langle j_{j \to i} \rangle = -\frac{i}{2} \left\langle \Psi_i^{\dagger} \check{t}_{ij} \Psi_j - \Psi_j^{\dagger} \check{t}_{ij}^{\dagger} \Psi_i \right\rangle \label{numformeq13}
 \end{eqnarray}
can be evaluated in terms of the correlation matrix elements involving the Green's function $\check{G}$ (Eq.\ \ref{numformeq9}) and self energy {\bf $\Gamma$} (Eq.\ \ref{numformeq8}) and is given by 
\begin{widetext} 
 \begin{eqnarray}
 	C_{x'\eta',x\eta}(t) 
 	&=& \langle \Psi^\dagger_{x'\eta'}(t) \Psi_{x\eta}(t)\rangle = \int \frac{d\omega}{2\pi} \sum_\lambda
 	\big[ \bG(\omega)\,\bGamma^\lambda\,\bG^\dagger(\omega) 
 	\big]_{x\eta,x'\eta'}\,
 	f^\lambda(\omega-\mu^\lambda),  \label{numformeq14}
 \end{eqnarray}
 \end{widetext} 
where $x,x'$ are site coordinates and $\eta,\eta'$ are indices for internal degrees of freedom. We have used Eq.\ \ref{numformeq13} to numerically compute the Josephson current in Sec.~\ref{sec:jr} where we sum over each bond (from left to right of the junction) over the width and get the total current $J(\varphi)$.

\bibliography{index_ref}

\begin{thebibliography}{46}%
\makeatletter
\providecommand \@ifxundefined [1]{%
 \@ifx{#1\undefined}
}%
\providecommand \@ifnum [1]{%
 \ifnum #1\expandafter \@firstoftwo
 \else \expandafter \@secondoftwo
 \fi
}%
\providecommand \@ifx [1]{%
 \ifx #1\expandafter \@firstoftwo
 \else \expandafter \@secondoftwo
 \fi
}%
\providecommand \natexlab [1]{#1}%
\providecommand \enquote  [1]{``#1''}%
\providecommand \bibnamefont  [1]{#1}%
\providecommand \bibfnamefont [1]{#1}%
\providecommand \citenamefont [1]{#1}%
\providecommand \href@noop [0]{\@secondoftwo}%
\providecommand \href [0]{\begingroup \@sanitize@url \@href}%
\providecommand \@href[1]{\@@startlink{#1}\@@href}%
\providecommand \@@href[1]{\endgroup#1\@@endlink}%
\providecommand \@sanitize@url [0]{\catcode `\\12\catcode `\$12\catcode
  `\&12\catcode `\#12\catcode `\^12\catcode `\_12\catcode `\%12\relax}%
\providecommand \@@startlink[1]{}%
\providecommand \@@endlink[0]{}%
\providecommand \url  [0]{\begingroup\@sanitize@url \@url }%
\providecommand \@url [1]{\endgroup\@href {#1}{\urlprefix }}%
\providecommand \urlprefix  [0]{URL }%
\providecommand \Eprint [0]{\href }%
\providecommand \doibase [0]{https://doi.org/}%
\providecommand \selectlanguage [0]{\@gobble}%
\providecommand \bibinfo  [0]{\@secondoftwo}%
\providecommand \bibfield  [0]{\@secondoftwo}%
\providecommand \translation [1]{[#1]}%
\providecommand \BibitemOpen [0]{}%
\providecommand \bibitemStop [0]{}%
\providecommand \bibitemNoStop [0]{.\EOS\space}%
\providecommand \EOS [0]{\spacefactor3000\relax}%
\providecommand \BibitemShut  [1]{\csname bibitem#1\endcsname}%
\let\auto@bib@innerbib\@empty
%</preamble>
\bibitem [{\citenamefont {Josephson}(1962)}]{joseph1}%
  \BibitemOpen
  \bibfield  {author} {\bibinfo {author} {\bibfnamefont {B.~D.}\ \bibnamefont
  {Josephson}},\ }\bibfield  {title} {\bibinfo {title} {Possible new effects in
  superconductive tunnelling},\ }\href
  {https://doi.org/10.1016/0031-9163(62)91369-0} {\bibfield  {journal}
  {\bibinfo  {journal} {Physics Letters}\ }\textbf {\bibinfo {volume} {1}},\
  \bibinfo {pages} {251} (\bibinfo {year} {1962})}\BibitemShut {NoStop}%
\bibitem [{\citenamefont {Golubov}\ \emph {et~al.}(2004)\citenamefont
  {Golubov}, \citenamefont {Kupriyanov},\ and\ \citenamefont
  {Il'ichev}}]{josephrev}%
  \BibitemOpen
  \bibfield  {author} {\bibinfo {author} {\bibfnamefont {A.~A.}\ \bibnamefont
  {Golubov}}, \bibinfo {author} {\bibfnamefont {M.~Y.}\ \bibnamefont
  {Kupriyanov}},\ and\ \bibinfo {author} {\bibfnamefont {E.}~\bibnamefont
  {Il'ichev}},\ }\bibfield  {title} {\bibinfo {title} {The current-phase
  relation in josephson junctions},\ }\href
  {https://doi.org/10.1103/RevModPhys.76.411} {\bibfield  {journal} {\bibinfo
  {journal} {Rev. Mod. Phys.}\ }\textbf {\bibinfo {volume} {76}},\ \bibinfo
  {pages} {411} (\bibinfo {year} {2004})}\BibitemShut {NoStop}%
\bibitem [{\citenamefont {Likharev}(1979)}]{likh1}%
  \BibitemOpen
  \bibfield  {author} {\bibinfo {author} {\bibfnamefont {K.~K.}\ \bibnamefont
  {Likharev}},\ }\bibfield  {title} {\bibinfo {title} {Superconducting weak
  links},\ }\href {https://doi.org/10.1103/RevModPhys.51.101} {\bibfield
  {journal} {\bibinfo  {journal} {Rev. Mod. Phys.}\ }\textbf {\bibinfo {volume}
  {51}},\ \bibinfo {pages} {101} (\bibinfo {year} {1979})}\BibitemShut
  {NoStop}%
\bibitem [{\citenamefont {Kwon}\ \emph {et~al.}(2004)\citenamefont {Kwon},
  \citenamefont {Sengupta},\ and\ \citenamefont {Yakovenko}}]{jos1}%
  \BibitemOpen
  \bibfield  {author} {\bibinfo {author} {\bibfnamefont {H.-J.}\ \bibnamefont
  {Kwon}}, \bibinfo {author} {\bibfnamefont {K.}~\bibnamefont {Sengupta}},\
  and\ \bibinfo {author} {\bibfnamefont {V.~M.}\ \bibnamefont {Yakovenko}},\
  }\bibfield  {title} {\bibinfo {title} {Fractional ac josephson effect in p-
  and d-wave superconductors},\ }\href
  {https://doi.org/10.1140/epjb/e2004-00066-4} {\bibfield  {journal} {\bibinfo
  {journal} {The European Physical Journal B - Condensed Matter and Complex
  Systems}\ }\textbf {\bibinfo {volume} {37}},\ \bibinfo {pages} {349}
  (\bibinfo {year} {2004})}\BibitemShut {NoStop}%
\bibitem [{\citenamefont {Kitaev}(2001)}]{kit1}%
  \BibitemOpen
  \bibfield  {author} {\bibinfo {author} {\bibfnamefont {A.~Y.}\ \bibnamefont
  {Kitaev}},\ }\bibfield  {title} {\bibinfo {title} {{Unpaired Majorana
  fermions in quantum wires}},\ }\href
  {https://doi.org/10.1070/1063-7869/44/10/S29} {\bibfield  {journal} {\bibinfo
   {journal} {Physics-Uspekhi}\ }\textbf {\bibinfo {volume} {44}},\ \bibinfo
  {pages} {131} (\bibinfo {year} {2001})}\BibitemShut {NoStop}%
\bibitem [{\citenamefont {Sticlet}\ \emph {et~al.}(2013)\citenamefont
  {Sticlet}, \citenamefont {Bena},\ and\ \citenamefont {Simon}}]{refparity}%
  \BibitemOpen
  \bibfield  {author} {\bibinfo {author} {\bibfnamefont {D.}~\bibnamefont
  {Sticlet}}, \bibinfo {author} {\bibfnamefont {C.}~\bibnamefont {Bena}},\ and\
  \bibinfo {author} {\bibfnamefont {P.}~\bibnamefont {Simon}},\ }\bibfield
  {title} {\bibinfo {title} {Josephson effect in superconducting wires
  supporting multiple majorana edge states},\ }\href
  {https://doi.org/10.1103/PhysRevB.87.104509} {\bibfield  {journal} {\bibinfo
  {journal} {Phys. Rev. B}\ }\textbf {\bibinfo {volume} {87}},\ \bibinfo
  {pages} {104509} (\bibinfo {year} {2013})}\BibitemShut {NoStop}%
\bibitem [{\citenamefont {J{\'e}rome}(2004)}]{tmtsfexp1}%
  \BibitemOpen
  \bibfield  {author} {\bibinfo {author} {\bibfnamefont {D.}~\bibnamefont
  {J{\'e}rome}},\ }\bibfield  {title} {\bibinfo {title} {{Organic conductors:
  From charge density wave TTF-TCNQ to superconducting (TMTSF)2PF6}},\ }\href
  {https://doi.org/10.1021/cr030652g} {\bibfield  {journal} {\bibinfo
  {journal} {{Chemical Reviews}}\ }\textbf {\bibinfo {volume} {104}},\ \bibinfo
  {pages} {5565} (\bibinfo {year} {2004})}\BibitemShut {NoStop}%
\bibitem [{\citenamefont {Parkin}\ \emph {et~al.}(1981)\citenamefont {Parkin},
  \citenamefont {Ribault}, \citenamefont {Jerome},\ and\ \citenamefont
  {Bechgaard}}]{tmtsfexp2}%
  \BibitemOpen
  \bibfield  {author} {\bibinfo {author} {\bibfnamefont {S.}~\bibnamefont
  {Parkin}}, \bibinfo {author} {\bibfnamefont {M.}~\bibnamefont {Ribault}},
  \bibinfo {author} {\bibfnamefont {D.}~\bibnamefont {Jerome}},\ and\ \bibinfo
  {author} {\bibfnamefont {K.}~\bibnamefont {Bechgaard}},\ }\bibfield  {title}
  {\bibinfo {title} {{Superconductivity in the family of organic salts based on
  the tetramethyltetraselenafulvalene (TMTSF) molecule: (TMTSF)2X (X=ClO4, PF6,
  AsF6, SbF6, TaF 6)}},\ }\href {https://doi.org/10.1088/0022-3719/14/34/011}
  {\bibfield  {journal} {\bibinfo  {journal} {{Journal of Physics C: Solid
  State Physics}}\ }\textbf {\bibinfo {volume} {14}},\ \bibinfo {pages} {5305}
  (\bibinfo {year} {1981})}\BibitemShut {NoStop}%
\bibitem [{\citenamefont {Ishiguro}\ \emph {et~al.}(1998)\citenamefont
  {Ishiguro}, \citenamefont {Yamaji},\ and\ \citenamefont {Saito}}]{ishi1}%
  \BibitemOpen
  \bibfield  {author} {\bibinfo {author} {\bibfnamefont {T.}~\bibnamefont
  {Ishiguro}}, \bibinfo {author} {\bibfnamefont {K.}~\bibnamefont {Yamaji}},\
  and\ \bibinfo {author} {\bibfnamefont {G.}~\bibnamefont {Saito}},\ }\bibinfo
  {title} {Organic conductors},\ in\ \href
  {https://doi.org/10.1007/978-3-642-58262-2_2} {\emph {\bibinfo {booktitle}
  {Organic Superconductors}}}\ (\bibinfo  {publisher} {Springer Berlin
  Heidelberg},\ \bibinfo {address} {Berlin, Heidelberg},\ \bibinfo {year}
  {1998})\ pp.\ \bibinfo {pages} {15--43}\BibitemShut {NoStop}%
\bibitem [{\citenamefont {Kwon}\ \emph {et~al.}(2003)\citenamefont {Kwon},
  \citenamefont {Yakovenko},\ and\ \citenamefont {Sengupta}}]{ks0}%
  \BibitemOpen
  \bibfield  {author} {\bibinfo {author} {\bibfnamefont {H.-J.}\ \bibnamefont
  {Kwon}}, \bibinfo {author} {\bibfnamefont {V.~M.}\ \bibnamefont
  {Yakovenko}},\ and\ \bibinfo {author} {\bibfnamefont {K.}~\bibnamefont
  {Sengupta}},\ }\bibfield  {title} {\bibinfo {title} {How to detect edge
  electron states in (tmtsf)2x and sr2ruo4 experimentally},\ }\href
  {https://doi.org/https://doi.org/10.1016/S0379-6779(02)00421-6} {\bibfield
  {journal} {\bibinfo  {journal} {Synthetic Metals}\ }\textbf {\bibinfo
  {volume} {133-134}},\ \bibinfo {pages} {27} (\bibinfo {year} {2003})},\
  \bibinfo {note} {proceedings of the Yamada Conference LVI. The Fourth
  International Symposium on Crystalline Organic Metals, Superconductors and
  Ferromagnets (ISCOM 2001).}\BibitemShut {Stop}%
\bibitem [{\citenamefont {Sengupta}\ \emph {et~al.}(2001)\citenamefont
  {Sengupta}, \citenamefont {\ifmmode \check{Z}\else
  \v{Z}\fi{}uti\ifmmode~\acute{c}\else \'{c}\fi{}}, \citenamefont {Kwon},
  \citenamefont {Yakovenko},\ and\ \citenamefont {Das~Sarma}}]{vmy0}%
  \BibitemOpen
  \bibfield  {author} {\bibinfo {author} {\bibfnamefont {K.}~\bibnamefont
  {Sengupta}}, \bibinfo {author} {\bibfnamefont {I.}~\bibnamefont {\ifmmode
  \check{Z}\else \v{Z}\fi{}uti\ifmmode~\acute{c}\else \'{c}\fi{}}}, \bibinfo
  {author} {\bibfnamefont {H.-J.}\ \bibnamefont {Kwon}}, \bibinfo {author}
  {\bibfnamefont {V.~M.}\ \bibnamefont {Yakovenko}},\ and\ \bibinfo {author}
  {\bibfnamefont {S.}~\bibnamefont {Das~Sarma}},\ }\bibfield  {title} {\bibinfo
  {title} {Midgap edge states and pairing symmetry of quasi-one-dimensional
  organic superconductors},\ }\href
  {https://doi.org/10.1103/PhysRevB.63.144531} {\bibfield  {journal} {\bibinfo
  {journal} {Phys. Rev. B}\ }\textbf {\bibinfo {volume} {63}},\ \bibinfo
  {pages} {144531} (\bibinfo {year} {2001})}\BibitemShut {NoStop}%
\bibitem [{\citenamefont {\ifmmode~\check{S}\else \v{S}\fi{}mejkal}\ \emph
  {et~al.}(2022{\natexlab{a}})\citenamefont {\ifmmode~\check{S}\else
  \v{S}\fi{}mejkal}, \citenamefont {Sinova},\ and\ \citenamefont
  {Jungwirth}}]{alt1}%
  \BibitemOpen
  \bibfield  {author} {\bibinfo {author} {\bibfnamefont {L.}~\bibnamefont
  {\ifmmode~\check{S}\else \v{S}\fi{}mejkal}}, \bibinfo {author} {\bibfnamefont
  {J.}~\bibnamefont {Sinova}},\ and\ \bibinfo {author} {\bibfnamefont
  {T.}~\bibnamefont {Jungwirth}},\ }\bibfield  {title} {\bibinfo {title}
  {Beyond conventional ferromagnetism and antiferromagnetism: A phase with
  nonrelativistic spin and crystal rotation symmetry},\ }\href
  {https://doi.org/10.1103/PhysRevX.12.031042} {\bibfield  {journal} {\bibinfo
  {journal} {Phys. Rev. X}\ }\textbf {\bibinfo {volume} {12}},\ \bibinfo
  {pages} {031042} (\bibinfo {year} {2022}{\natexlab{a}})}\BibitemShut
  {NoStop}%
\bibitem [{\citenamefont {\ifmmode~\check{S}\else \v{S}\fi{}mejkal}\ \emph
  {et~al.}(2022{\natexlab{b}})\citenamefont {\ifmmode~\check{S}\else
  \v{S}\fi{}mejkal}, \citenamefont {Sinova},\ and\ \citenamefont
  {Jungwirth}}]{alt2}%
  \BibitemOpen
  \bibfield  {author} {\bibinfo {author} {\bibfnamefont {L.}~\bibnamefont
  {\ifmmode~\check{S}\else \v{S}\fi{}mejkal}}, \bibinfo {author} {\bibfnamefont
  {J.}~\bibnamefont {Sinova}},\ and\ \bibinfo {author} {\bibfnamefont
  {T.}~\bibnamefont {Jungwirth}},\ }\bibfield  {title} {\bibinfo {title}
  {Emerging research landscape of altermagnetism},\ }\href
  {https://doi.org/10.1103/PhysRevX.12.040501} {\bibfield  {journal} {\bibinfo
  {journal} {Phys. Rev. X}\ }\textbf {\bibinfo {volume} {12}},\ \bibinfo
  {pages} {040501} (\bibinfo {year} {2022}{\natexlab{b}})}\BibitemShut
  {NoStop}%
\bibitem [{\citenamefont {Brekke}\ \emph
  {et~al.}(2023{\natexlab{a}})\citenamefont {Brekke}, \citenamefont {Brataas},\
  and\ \citenamefont {Sudb\o{}}}]{alt3}%
  \BibitemOpen
  \bibfield  {author} {\bibinfo {author} {\bibfnamefont {B.}~\bibnamefont
  {Brekke}}, \bibinfo {author} {\bibfnamefont {A.}~\bibnamefont {Brataas}},\
  and\ \bibinfo {author} {\bibfnamefont {A.}~\bibnamefont {Sudb\o{}}},\
  }\bibfield  {title} {\bibinfo {title} {Two-dimensional altermagnets:
  Superconductivity in a minimal microscopic model},\ }\href
  {https://doi.org/10.1103/PhysRevB.108.224421} {\bibfield  {journal} {\bibinfo
   {journal} {Phys. Rev. B}\ }\textbf {\bibinfo {volume} {108}},\ \bibinfo
  {pages} {224421} (\bibinfo {year} {2023}{\natexlab{a}})}\BibitemShut
  {NoStop}%
\bibitem [{\citenamefont {Krempask{\~A}{\textonehalf}}\ \emph
  {et~al.}(2024)\citenamefont {Krempask{\~A}{\textonehalf}}, \citenamefont
  {{\AA} mejkal}, \citenamefont {D{\^a}{\texteuro}{\texttrademark}Souza},
  \citenamefont {Hajlaoui}, \citenamefont {Springholz}, \citenamefont
  {Uhl{\~A}­{\AA}{\texttrademark}ov{\~A}{\textexclamdown}}, \citenamefont
  {Alarab}, \citenamefont {Constantinou}, \citenamefont {Strocov},
  \citenamefont {Usanov}, \citenamefont {Pudelko}, \citenamefont
  {Gonz{\~A}{\textexclamdown}lez-Hern{\~A}{\textexclamdown}ndez}, \citenamefont
  {Birk~Hellenes}, \citenamefont {Jansa}, \citenamefont
  {Reichlov{\~A}{\textexclamdown}}, \citenamefont
  {{\AA} ob{\~A}{\textexclamdown}{\AA}ˆ}, \citenamefont
  {Gonzalez~Betancourt}, \citenamefont {Wadley}, \citenamefont {Sinova},
  \citenamefont {Kriegner}, \citenamefont {Min{\~A}{\textexclamdown}r},
  \citenamefont {Dil},\ and\ \citenamefont {Jungwirth}}]{alt4}%
  \BibitemOpen
  \bibfield  {author} {\bibinfo {author} {\bibfnamefont {J.}~\bibnamefont
  {Krempask{\~A}{\textonehalf}}}, \bibinfo {author} {\bibfnamefont
  {L.}~\bibnamefont {{\AA} mejkal}}, \bibinfo {author} {\bibfnamefont {S.~W.}\
  \bibnamefont {D{\^a}{\texteuro}{\texttrademark}Souza}}, \bibinfo {author}
  {\bibfnamefont {M.}~\bibnamefont {Hajlaoui}}, \bibinfo {author}
  {\bibfnamefont {G.}~\bibnamefont {Springholz}}, \bibinfo {author}
  {\bibfnamefont {K.}~\bibnamefont
  {Uhl{\~A}­{\AA}{\texttrademark}ov{\~A}{\textexclamdown}}}, \bibinfo {author}
  {\bibfnamefont {F.}~\bibnamefont {Alarab}}, \bibinfo {author} {\bibfnamefont
  {P.~C.}\ \bibnamefont {Constantinou}}, \bibinfo {author} {\bibfnamefont
  {V.}~\bibnamefont {Strocov}}, \bibinfo {author} {\bibfnamefont
  {D.}~\bibnamefont {Usanov}}, \bibinfo {author} {\bibfnamefont {W.~R.}\
  \bibnamefont {Pudelko}}, \bibinfo {author} {\bibfnamefont {R.}~\bibnamefont
  {Gonz{\~A}{\textexclamdown}lez-Hern{\~A}{\textexclamdown}ndez}}, \bibinfo
  {author} {\bibfnamefont {A.}~\bibnamefont {Birk~Hellenes}}, \bibinfo {author}
  {\bibfnamefont {Z.}~\bibnamefont {Jansa}}, \bibinfo {author} {\bibfnamefont
  {H.}~\bibnamefont {Reichlov{\~A}{\textexclamdown}}}, \bibinfo {author}
  {\bibfnamefont {Z.}~\bibnamefont {{\AA} ob{\~A}{\textexclamdown}{\AA}ˆ}},
  \bibinfo {author} {\bibfnamefont {R.~D.}\ \bibnamefont
  {Gonzalez~Betancourt}}, \bibinfo {author} {\bibfnamefont {P.}~\bibnamefont
  {Wadley}}, \bibinfo {author} {\bibfnamefont {J.}~\bibnamefont {Sinova}},
  \bibinfo {author} {\bibfnamefont {D.}~\bibnamefont {Kriegner}}, \bibinfo
  {author} {\bibfnamefont {J.}~\bibnamefont {Min{\~A}{\textexclamdown}r}},
  \bibinfo {author} {\bibfnamefont {J.~H.}\ \bibnamefont {Dil}},\ and\ \bibinfo
  {author} {\bibfnamefont {T.}~\bibnamefont {Jungwirth}},\ }\bibfield  {title}
  {\bibinfo {title} {Altermagnetic lifting of kramers spin degeneracy},\ }\href
  {https://doi.org/10.1038/s41586-023-06907-7} {\bibfield  {journal} {\bibinfo
  {journal} {Nature}\ }\textbf {\bibinfo {volume} {626}},\ \bibinfo {pages}
  {517} (\bibinfo {year} {2024})}\BibitemShut {NoStop}%
\bibitem [{\citenamefont {Lee}\ \emph {et~al.}(2024)\citenamefont {Lee},
  \citenamefont {Lee}, \citenamefont {Jung}, \citenamefont {Jung},
  \citenamefont {Kim}, \citenamefont {Lee}, \citenamefont {Seok}, \citenamefont
  {Kim}, \citenamefont {Park}, \citenamefont {\ifmmode~\check{S}\else
  \v{S}\fi{}mejkal}, \citenamefont {Kang},\ and\ \citenamefont {Kim}}]{alt5}%
  \BibitemOpen
  \bibfield  {author} {\bibinfo {author} {\bibfnamefont {S.}~\bibnamefont
  {Lee}}, \bibinfo {author} {\bibfnamefont {S.}~\bibnamefont {Lee}}, \bibinfo
  {author} {\bibfnamefont {S.}~\bibnamefont {Jung}}, \bibinfo {author}
  {\bibfnamefont {J.}~\bibnamefont {Jung}}, \bibinfo {author} {\bibfnamefont
  {D.}~\bibnamefont {Kim}}, \bibinfo {author} {\bibfnamefont {Y.}~\bibnamefont
  {Lee}}, \bibinfo {author} {\bibfnamefont {B.}~\bibnamefont {Seok}}, \bibinfo
  {author} {\bibfnamefont {J.}~\bibnamefont {Kim}}, \bibinfo {author}
  {\bibfnamefont {B.~G.}\ \bibnamefont {Park}}, \bibinfo {author}
  {\bibfnamefont {L.}~\bibnamefont {\ifmmode~\check{S}\else \v{S}\fi{}mejkal}},
  \bibinfo {author} {\bibfnamefont {C.-J.}\ \bibnamefont {Kang}},\ and\
  \bibinfo {author} {\bibfnamefont {C.}~\bibnamefont {Kim}},\ }\bibfield
  {title} {\bibinfo {title} {Broken kramers degeneracy in altermagnetic mnte},\
  }\href {https://doi.org/10.1103/PhysRevLett.132.036702} {\bibfield  {journal}
  {\bibinfo  {journal} {Phys. Rev. Lett.}\ }\textbf {\bibinfo {volume} {132}},\
  \bibinfo {pages} {036702} (\bibinfo {year} {2024})}\BibitemShut {NoStop}%
\bibitem [{\citenamefont {Zhu}\ \emph {et~al.}(2023)\citenamefont {Zhu},
  \citenamefont {Zhuang}, \citenamefont {Wu},\ and\ \citenamefont
  {Yan}}]{alt6}%
  \BibitemOpen
  \bibfield  {author} {\bibinfo {author} {\bibfnamefont {D.}~\bibnamefont
  {Zhu}}, \bibinfo {author} {\bibfnamefont {Z.-Y.}\ \bibnamefont {Zhuang}},
  \bibinfo {author} {\bibfnamefont {Z.}~\bibnamefont {Wu}},\ and\ \bibinfo
  {author} {\bibfnamefont {Z.}~\bibnamefont {Yan}},\ }\bibfield  {title}
  {\bibinfo {title} {Topological superconductivity in two-dimensional
  altermagnetic metals},\ }\href {https://doi.org/10.1103/PhysRevB.108.184505}
  {\bibfield  {journal} {\bibinfo  {journal} {Phys. Rev. B}\ }\textbf {\bibinfo
  {volume} {108}},\ \bibinfo {pages} {184505} (\bibinfo {year}
  {2023})}\BibitemShut {NoStop}%
\bibitem [{\citenamefont {Bose}\ \emph
  {et~al.}(2024{\natexlab{a}})\citenamefont {Bose}, \citenamefont {Vadnais},\
  and\ \citenamefont {Paramekanti}}]{alt7}%
  \BibitemOpen
  \bibfield  {author} {\bibinfo {author} {\bibfnamefont {A.}~\bibnamefont
  {Bose}}, \bibinfo {author} {\bibfnamefont {S.}~\bibnamefont {Vadnais}},\ and\
  \bibinfo {author} {\bibfnamefont {A.}~\bibnamefont {Paramekanti}},\
  }\bibfield  {title} {\bibinfo {title} {Altermagnetism and superconductivity
  in a multiorbital $t\ensuremath{-}j$ model},\ }\href
  {https://doi.org/10.1103/PhysRevB.110.205120} {\bibfield  {journal} {\bibinfo
   {journal} {Phys. Rev. B}\ }\textbf {\bibinfo {volume} {110}},\ \bibinfo
  {pages} {205120} (\bibinfo {year} {2024}{\natexlab{a}})}\BibitemShut
  {NoStop}%
\bibitem [{\citenamefont {Bose}\ \emph
  {et~al.}(2024{\natexlab{b}})\citenamefont {Bose}, \citenamefont {Vadnais},\
  and\ \citenamefont {Paramekanti}}]{alt8}%
  \BibitemOpen
  \bibfield  {author} {\bibinfo {author} {\bibfnamefont {A.}~\bibnamefont
  {Bose}}, \bibinfo {author} {\bibfnamefont {S.}~\bibnamefont {Vadnais}},\ and\
  \bibinfo {author} {\bibfnamefont {A.}~\bibnamefont {Paramekanti}},\
  }\bibfield  {title} {\bibinfo {title} {Altermagnetism and superconductivity
  in a multiorbital $t\ensuremath{-}j$ model},\ }\href
  {https://doi.org/10.1103/PhysRevB.110.205120} {\bibfield  {journal} {\bibinfo
   {journal} {Phys. Rev. B}\ }\textbf {\bibinfo {volume} {110}},\ \bibinfo
  {pages} {205120} (\bibinfo {year} {2024}{\natexlab{b}})}\BibitemShut
  {NoStop}%
\bibitem [{\citenamefont {McClarty}\ and\ \citenamefont {Rau}(2024)}]{alt9}%
  \BibitemOpen
  \bibfield  {author} {\bibinfo {author} {\bibfnamefont {P.~A.}\ \bibnamefont
  {McClarty}}\ and\ \bibinfo {author} {\bibfnamefont {J.~G.}\ \bibnamefont
  {Rau}},\ }\bibfield  {title} {\bibinfo {title} {Landau theory of
  altermagnetism},\ }\href {https://doi.org/10.1103/PhysRevLett.132.176702}
  {\bibfield  {journal} {\bibinfo  {journal} {Phys. Rev. Lett.}\ }\textbf
  {\bibinfo {volume} {132}},\ \bibinfo {pages} {176702} (\bibinfo {year}
  {2024})}\BibitemShut {NoStop}%
\bibitem [{\citenamefont {Ouassou}\ \emph {et~al.}(2023)\citenamefont
  {Ouassou}, \citenamefont {Brataas},\ and\ \citenamefont {Linder}}]{alt10}%
  \BibitemOpen
  \bibfield  {author} {\bibinfo {author} {\bibfnamefont {J.~A.}\ \bibnamefont
  {Ouassou}}, \bibinfo {author} {\bibfnamefont {A.}~\bibnamefont {Brataas}},\
  and\ \bibinfo {author} {\bibfnamefont {J.}~\bibnamefont {Linder}},\
  }\bibfield  {title} {\bibinfo {title} {dc josephson effect in altermagnets},\
  }\href {https://doi.org/10.1103/PhysRevLett.131.076003} {\bibfield  {journal}
  {\bibinfo  {journal} {Phys. Rev. Lett.}\ }\textbf {\bibinfo {volume} {131}},\
  \bibinfo {pages} {076003} (\bibinfo {year} {2023})}\BibitemShut {NoStop}%
\bibitem [{\citenamefont {Zhang}\ \emph {et~al.}(2024)\citenamefont {Zhang},
  \citenamefont {Hu},\ and\ \citenamefont {Neupert}}]{alt11}%
  \BibitemOpen
  \bibfield  {author} {\bibinfo {author} {\bibfnamefont {S.-B.}\ \bibnamefont
  {Zhang}}, \bibinfo {author} {\bibfnamefont {L.-H.}\ \bibnamefont {Hu}},\ and\
  \bibinfo {author} {\bibfnamefont {T.}~\bibnamefont {Neupert}},\ }\bibfield
  {title} {\bibinfo {title} {Finite-momentum cooper pairing in proximitized
  altermagnets},\ }\href {https://doi.org/10.1038/s41467-024-45951-3}
  {\bibfield  {journal} {\bibinfo  {journal} {Nature Communications}\ }\textbf
  {\bibinfo {volume} {15}},\ \bibinfo {pages} {1801} (\bibinfo {year}
  {2024})}\BibitemShut {NoStop}%
\bibitem [{\citenamefont {Ghorashi}\ \emph {et~al.}(2024)\citenamefont
  {Ghorashi}, \citenamefont {Hughes},\ and\ \citenamefont {Cano}}]{alt12}%
  \BibitemOpen
  \bibfield  {author} {\bibinfo {author} {\bibfnamefont {S.~A.~A.}\
  \bibnamefont {Ghorashi}}, \bibinfo {author} {\bibfnamefont {T.~L.}\
  \bibnamefont {Hughes}},\ and\ \bibinfo {author} {\bibfnamefont
  {J.}~\bibnamefont {Cano}},\ }\bibfield  {title} {\bibinfo {title}
  {Altermagnetic routes to majorana modes in zero net magnetization},\ }\href
  {https://doi.org/10.1103/PhysRevLett.133.106601} {\bibfield  {journal}
  {\bibinfo  {journal} {Phys. Rev. Lett.}\ }\textbf {\bibinfo {volume} {133}},\
  \bibinfo {pages} {106601} (\bibinfo {year} {2024})}\BibitemShut {NoStop}%
\bibitem [{\citenamefont {Liu}\ \emph {et~al.}(2025)\citenamefont {Liu},
  \citenamefont {Hu},\ and\ \citenamefont {Liu}}]{revsup}%
  \BibitemOpen
  \bibfield  {author} {\bibinfo {author} {\bibfnamefont {Z.}~\bibnamefont
  {Liu}}, \bibinfo {author} {\bibfnamefont {H.}~\bibnamefont {Hu}},\ and\
  \bibinfo {author} {\bibfnamefont {X.-J.}\ \bibnamefont {Liu}},\ }\href
  {https://arxiv.org/abs/2510.09170} {\bibinfo {title} {Altermagnetism and
  superconductivity: A short historical review}} (\bibinfo {year} {2025}),\
  \Eprint {https://arxiv.org/abs/2510.09170} {arXiv:2510.09170
  [cond-mat.supr-con]} \BibitemShut {NoStop}%
\bibitem [{\citenamefont {Brekke}\ \emph
  {et~al.}(2023{\natexlab{b}})\citenamefont {Brekke}, \citenamefont {Brataas},\
  and\ \citenamefont {Sudb\o{}}}]{sudbo1}%
  \BibitemOpen
  \bibfield  {author} {\bibinfo {author} {\bibfnamefont {B.}~\bibnamefont
  {Brekke}}, \bibinfo {author} {\bibfnamefont {A.}~\bibnamefont {Brataas}},\
  and\ \bibinfo {author} {\bibfnamefont {A.}~\bibnamefont {Sudb\o{}}},\
  }\bibfield  {title} {\bibinfo {title} {Two-dimensional altermagnets:
  Superconductivity in a minimal microscopic model},\ }\href
  {https://doi.org/10.1103/PhysRevB.108.224421} {\bibfield  {journal} {\bibinfo
   {journal} {Phys. Rev. B}\ }\textbf {\bibinfo {volume} {108}},\ \bibinfo
  {pages} {224421} (\bibinfo {year} {2023}{\natexlab{b}})}\BibitemShut
  {NoStop}%
\bibitem [{\citenamefont {Leraand}\ \emph {et~al.}(2025)\citenamefont
  {Leraand}, \citenamefont {M\ae{}land},\ and\ \citenamefont
  {Sudb\o{}}}]{sudbo2}%
  \BibitemOpen
  \bibfield  {author} {\bibinfo {author} {\bibfnamefont {K.}~\bibnamefont
  {Leraand}}, \bibinfo {author} {\bibfnamefont {K.}~\bibnamefont
  {M\ae{}land}},\ and\ \bibinfo {author} {\bibfnamefont {A.}~\bibnamefont
  {Sudb\o{}}},\ }\bibfield  {title} {\bibinfo {title} {Phonon-mediated
  spin-polarized superconductivity in altermagnets},\ }\href
  {https://doi.org/10.1103/g4dl-1ff2} {\bibfield  {journal} {\bibinfo
  {journal} {Phys. Rev. B}\ }\textbf {\bibinfo {volume} {112}},\ \bibinfo
  {pages} {104510} (\bibinfo {year} {2025})}\BibitemShut {NoStop}%
\bibitem [{\citenamefont {Chakraborty}\ and\ \citenamefont
  {Black-Schaffer}(2024)}]{sup1}%
  \BibitemOpen
  \bibfield  {author} {\bibinfo {author} {\bibfnamefont {D.}~\bibnamefont
  {Chakraborty}}\ and\ \bibinfo {author} {\bibfnamefont {A.~M.}\ \bibnamefont
  {Black-Schaffer}},\ }\bibfield  {title} {\bibinfo {title} {Zero-field
  finite-momentum and field-induced superconductivity in altermagnets},\ }\href
  {https://doi.org/10.1103/PhysRevB.110.L060508} {\bibfield  {journal}
  {\bibinfo  {journal} {Phys. Rev. B}\ }\textbf {\bibinfo {volume} {110}},\
  \bibinfo {pages} {L060508} (\bibinfo {year} {2024})}\BibitemShut {NoStop}%
\bibitem [{\citenamefont {de~Carvalho}\ and\ \citenamefont
  {Freire}(2024)}]{sup2}%
  \BibitemOpen
  \bibfield  {author} {\bibinfo {author} {\bibfnamefont {V.~S.}\ \bibnamefont
  {de~Carvalho}}\ and\ \bibinfo {author} {\bibfnamefont {H.}~\bibnamefont
  {Freire}},\ }\bibfield  {title} {\bibinfo {title} {Unconventional
  superconductivity in altermagnets with spin-orbit coupling},\ }\href
  {https://doi.org/10.1103/PhysRevB.110.L220503} {\bibfield  {journal}
  {\bibinfo  {journal} {Phys. Rev. B}\ }\textbf {\bibinfo {volume} {110}},\
  \bibinfo {pages} {L220503} (\bibinfo {year} {2024})}\BibitemShut {NoStop}%
\bibitem [{\citenamefont {Hong}\ \emph {et~al.}(2025)\citenamefont {Hong},
  \citenamefont {Park},\ and\ \citenamefont {Kim}}]{sup3}%
  \BibitemOpen
  \bibfield  {author} {\bibinfo {author} {\bibfnamefont {S.}~\bibnamefont
  {Hong}}, \bibinfo {author} {\bibfnamefont {M.~J.}\ \bibnamefont {Park}},\
  and\ \bibinfo {author} {\bibfnamefont {K.-M.}\ \bibnamefont {Kim}},\
  }\bibfield  {title} {\bibinfo {title} {Unconventional $p$-wave and
  finite-momentum superconductivity induced by altermagnetism through the
  formation of bogoliubov fermi surface},\ }\href
  {https://doi.org/10.1103/PhysRevB.111.054501} {\bibfield  {journal} {\bibinfo
   {journal} {Phys. Rev. B}\ }\textbf {\bibinfo {volume} {111}},\ \bibinfo
  {pages} {054501} (\bibinfo {year} {2025})}\BibitemShut {NoStop}%
\bibitem [{\citenamefont {Lu}\ \emph {et~al.}(2024)\citenamefont {Lu},
  \citenamefont {Maeda}, \citenamefont {Ito}, \citenamefont {Yada},\ and\
  \citenamefont {Tanaka}}]{sct1}%
  \BibitemOpen
  \bibfield  {author} {\bibinfo {author} {\bibfnamefont {B.}~\bibnamefont
  {Lu}}, \bibinfo {author} {\bibfnamefont {K.}~\bibnamefont {Maeda}}, \bibinfo
  {author} {\bibfnamefont {H.}~\bibnamefont {Ito}}, \bibinfo {author}
  {\bibfnamefont {K.}~\bibnamefont {Yada}},\ and\ \bibinfo {author}
  {\bibfnamefont {Y.}~\bibnamefont {Tanaka}},\ }\bibfield  {title} {\bibinfo
  {title} {$\ensuremath{\varphi}$ josephson junction induced by
  altermagnetism},\ }\href {https://doi.org/10.1103/PhysRevLett.133.226002}
  {\bibfield  {journal} {\bibinfo  {journal} {Phys. Rev. Lett.}\ }\textbf
  {\bibinfo {volume} {133}},\ \bibinfo {pages} {226002} (\bibinfo {year}
  {2024})}\BibitemShut {NoStop}%
\bibitem [{\citenamefont {Beenakker}\ and\ \citenamefont
  {Vakhtel}(2023)}]{sct2}%
  \BibitemOpen
  \bibfield  {author} {\bibinfo {author} {\bibfnamefont {C.~W.~J.}\
  \bibnamefont {Beenakker}}\ and\ \bibinfo {author} {\bibfnamefont
  {T.}~\bibnamefont {Vakhtel}},\ }\bibfield  {title} {\bibinfo {title}
  {Phase-shifted andreev levels in an altermagnet josephson junction},\ }\href
  {https://doi.org/10.1103/PhysRevB.108.075425} {\bibfield  {journal} {\bibinfo
   {journal} {Phys. Rev. B}\ }\textbf {\bibinfo {volume} {108}},\ \bibinfo
  {pages} {075425} (\bibinfo {year} {2023})}\BibitemShut {NoStop}%
\bibitem [{\citenamefont {Fukaya}\ \emph {et~al.}(2025)\citenamefont {Fukaya},
  \citenamefont {Lu}, \citenamefont {Yada}, \citenamefont {Tanaka},\ and\
  \citenamefont {Cayao}}]{sct3}%
  \BibitemOpen
  \bibfield  {author} {\bibinfo {author} {\bibfnamefont {Y.}~\bibnamefont
  {Fukaya}}, \bibinfo {author} {\bibfnamefont {B.}~\bibnamefont {Lu}}, \bibinfo
  {author} {\bibfnamefont {K.}~\bibnamefont {Yada}}, \bibinfo {author}
  {\bibfnamefont {Y.}~\bibnamefont {Tanaka}},\ and\ \bibinfo {author}
  {\bibfnamefont {J.}~\bibnamefont {Cayao}},\ }\bibfield  {title} {\bibinfo
  {title} {Superconducting phenomena in systems with unconventional magnets},\
  }\href {https://doi.org/10.1088/1361-648X/adf1cf} {\bibfield  {journal}
  {\bibinfo  {journal} {Journal of Physics: Condensed Matter}\ }\textbf
  {\bibinfo {volume} {37}},\ \bibinfo {pages} {313003} (\bibinfo {year}
  {2025})}\BibitemShut {NoStop}%
\bibitem [{\citenamefont {Sun}\ \emph {et~al.}(2023)\citenamefont {Sun},
  \citenamefont {Brataas},\ and\ \citenamefont {Linder}}]{sct4}%
  \BibitemOpen
  \bibfield  {author} {\bibinfo {author} {\bibfnamefont {C.}~\bibnamefont
  {Sun}}, \bibinfo {author} {\bibfnamefont {A.}~\bibnamefont {Brataas}},\ and\
  \bibinfo {author} {\bibfnamefont {J.}~\bibnamefont {Linder}},\ }\bibfield
  {title} {\bibinfo {title} {Andreev reflection in altermagnets},\ }\href
  {https://doi.org/10.1103/PhysRevB.108.054511} {\bibfield  {journal} {\bibinfo
   {journal} {Phys. Rev. B}\ }\textbf {\bibinfo {volume} {108}},\ \bibinfo
  {pages} {054511} (\bibinfo {year} {2023})}\BibitemShut {NoStop}%
\bibitem [{\citenamefont {Alipourzadeh}\ and\ \citenamefont
  {Hajati}(2025)}]{sct5}%
  \BibitemOpen
  \bibfield  {author} {\bibinfo {author} {\bibfnamefont {M.}~\bibnamefont
  {Alipourzadeh}}\ and\ \bibinfo {author} {\bibfnamefont {Y.}~\bibnamefont
  {Hajati}},\ }\bibfield  {title} {\bibinfo {title} {Andreev bound states and
  supercurrent in an unconventional superconductor-altermagnet josephson
  junction},\ }\href {https://doi.org/10.1103/mj4b-2fnr} {\bibfield  {journal}
  {\bibinfo  {journal} {Phys. Rev. B}\ }\textbf {\bibinfo {volume} {111}},\
  \bibinfo {pages} {214515} (\bibinfo {year} {2025})}\BibitemShut {NoStop}%
\bibitem [{\citenamefont {Papaj}(2023)}]{sct6}%
  \BibitemOpen
  \bibfield  {author} {\bibinfo {author} {\bibfnamefont {M.}~\bibnamefont
  {Papaj}},\ }\bibfield  {title} {\bibinfo {title} {Andreev reflection at the
  altermagnet-superconductor interface},\ }\href
  {https://doi.org/10.1103/PhysRevB.108.L060508} {\bibfield  {journal}
  {\bibinfo  {journal} {Phys. Rev. B}\ }\textbf {\bibinfo {volume} {108}},\
  \bibinfo {pages} {L060508} (\bibinfo {year} {2023})}\BibitemShut {NoStop}%
\bibitem [{\citenamefont {Junker}(1996)}]{susyref1}%
  \BibitemOpen
  \bibfield  {author} {\bibinfo {author} {\bibfnamefont {G.}~\bibnamefont
  {Junker}},\ }\bibinfo {title} {Supersymmetric quantum mechanics},\ in\ \href
  {https://doi.org/10.1007/978-3-642-61194-0_2} {\emph {\bibinfo {booktitle}
  {Supersymmetric Methods in Quantum and Statistical Physics}}}\ (\bibinfo
  {publisher} {Springer Berlin Heidelberg},\ \bibinfo {address} {Berlin,
  Heidelberg},\ \bibinfo {year} {1996})\ pp.\ \bibinfo {pages}
  {7--19}\BibitemShut {NoStop}%
\bibitem [{\citenamefont {Junker}(2019)}]{vic1}%
  \BibitemOpen
  \bibfield  {author} {\bibinfo {author} {\bibfnamefont {G.}~\bibnamefont
  {Junker}},\ }\href {https://doi.org/10.1088/2053-2563/aae6d5} {\emph
  {\bibinfo {title} {Supersymmetric Methods in Quantum, Statistical and Solid
  State Physics}}},\ 2053-2563\ (\bibinfo  {publisher} {IOP Publishing},\
  \bibinfo {year} {2019})\BibitemShut {NoStop}%
\bibitem [{\citenamefont {Witten}(2000)}]{index1}%
  \BibitemOpen
  \bibfield  {author} {\bibinfo {author} {\bibfnamefont {E.}~\bibnamefont
  {Witten}},\ }\href {https://arxiv.org/abs/hep-th/0006010} {\bibinfo {title}
  {Supersymmetric index in four-dimensional gauge theories}} (\bibinfo {year}
  {2000}),\ \Eprint {https://arxiv.org/abs/hep-th/0006010}
  {arXiv:hep-th/0006010 [hep-th]} \BibitemShut {NoStop}%
\bibitem [{\citenamefont {Brazovski\u{i}}(1980)}]{braz1}%
  \BibitemOpen
  \bibfield  {author} {\bibinfo {author} {\bibfnamefont {S.~A.}\ \bibnamefont
  {Brazovski\u{i}}},\ }\bibfield  {title} {\bibinfo {title} {Self-localized
  excitations in the {Peierls-Fr\"ohlich} state},\ }\href
  {http://jetp.ras.ru/cgi-bin/e/index/e/51/2/p342?a=list} {\bibfield  {journal}
  {\bibinfo  {journal} {JETP}\ }\textbf {\bibinfo {volume} {51}},\ \bibinfo
  {pages} {342} (\bibinfo {year} {1980})}\BibitemShut {NoStop}%
\bibitem [{\citenamefont {Sengupta}\ \emph {et~al.}(2002)\citenamefont
  {Sengupta}, \citenamefont {Kwon},\ and\ \citenamefont
  {M.~Yakovenko}}]{ruth1}%
  \BibitemOpen
  \bibfield  {author} {\bibinfo {author} {\bibfnamefont {K.}~\bibnamefont
  {Sengupta}}, \bibinfo {author} {\bibfnamefont {H.-J.}\ \bibnamefont {Kwon}},\
  and\ \bibinfo {author} {\bibfnamefont {V.}~\bibnamefont {M.~Yakovenko}},\
  }\bibfield  {title} {\bibinfo {title} {Edge states and determination of
  pairing symmetry in superconducting ${\mathrm{sr}}_{2}{\mathrm{ruo}}_{4}$},\
  }\href {https://doi.org/10.1103/PhysRevB.65.104504} {\bibfield  {journal}
  {\bibinfo  {journal} {Phys. Rev. B}\ }\textbf {\bibinfo {volume} {65}},\
  \bibinfo {pages} {104504} (\bibinfo {year} {2002})}\BibitemShut {NoStop}%
\bibitem [{\citenamefont {Jiang}\ \emph {et~al.}(2011)\citenamefont {Jiang},
  \citenamefont {Pekker}, \citenamefont {Alicea}, \citenamefont {Refael},
  \citenamefont {Oreg},\ and\ \citenamefont {von
  Oppen}}]{jiang2011unconventional}%
  \BibitemOpen
  \bibfield  {author} {\bibinfo {author} {\bibfnamefont {L.}~\bibnamefont
  {Jiang}}, \bibinfo {author} {\bibfnamefont {D.}~\bibnamefont {Pekker}},
  \bibinfo {author} {\bibfnamefont {J.}~\bibnamefont {Alicea}}, \bibinfo
  {author} {\bibfnamefont {G.}~\bibnamefont {Refael}}, \bibinfo {author}
  {\bibfnamefont {Y.}~\bibnamefont {Oreg}},\ and\ \bibinfo {author}
  {\bibfnamefont {F.}~\bibnamefont {von Oppen}},\ }\bibfield  {title} {\bibinfo
  {title} {Unconventional josephson signatures of majorana bound states},\
  }\href@noop {} {\bibfield  {journal} {\bibinfo  {journal} {Physical review
  letters}\ }\textbf {\bibinfo {volume} {107}},\ \bibinfo {pages} {236401}
  (\bibinfo {year} {2011})}\BibitemShut {NoStop}%
\bibitem [{\citenamefont {Kumari}\ \emph {et~al.}(2024)\citenamefont {Kumari},
  \citenamefont {Seradjeh},\ and\ \citenamefont {Kundu}}]{kumari2024josephson}%
  \BibitemOpen
  \bibfield  {author} {\bibinfo {author} {\bibfnamefont {R.}~\bibnamefont
  {Kumari}}, \bibinfo {author} {\bibfnamefont {B.}~\bibnamefont {Seradjeh}},\
  and\ \bibinfo {author} {\bibfnamefont {A.}~\bibnamefont {Kundu}},\ }\bibfield
   {title} {\bibinfo {title} {Josephson-current signatures of unpaired floquet
  majorana fermions},\ }\href@noop {} {\bibfield  {journal} {\bibinfo
  {journal} {Physical Review Letters}\ }\textbf {\bibinfo {volume} {133}},\
  \bibinfo {pages} {196601} (\bibinfo {year} {2024})}\BibitemShut {NoStop}%
\bibitem [{\citenamefont {Shapiro}(1963)}]{shap1}%
  \BibitemOpen
  \bibfield  {author} {\bibinfo {author} {\bibfnamefont {S.}~\bibnamefont
  {Shapiro}},\ }\bibfield  {title} {\bibinfo {title} {Josephson currents in
  superconducting tunneling: The effect of microwaves and other observations},\
  }\href {https://doi.org/10.1103/PhysRevLett.11.80} {\bibfield  {journal}
  {\bibinfo  {journal} {Phys. Rev. Lett.}\ }\textbf {\bibinfo {volume} {11}},\
  \bibinfo {pages} {80} (\bibinfo {year} {1963})}\BibitemShut {NoStop}%
\bibitem [{\citenamefont {Rokhinson}\ \emph {et~al.}(2012)\citenamefont
  {Rokhinson}, \citenamefont {Liu},\ and\ \citenamefont {Furdyna}}]{shap2}%
  \BibitemOpen
  \bibfield  {author} {\bibinfo {author} {\bibfnamefont {L.~P.}\ \bibnamefont
  {Rokhinson}}, \bibinfo {author} {\bibfnamefont {X.}~\bibnamefont {Liu}},\
  and\ \bibinfo {author} {\bibfnamefont {J.~K.}\ \bibnamefont {Furdyna}},\
  }\bibfield  {title} {\bibinfo {title} {The fractional a.c. josephson effect
  in a semiconductor--superconductor nanowire as a signature of majorana
  particles},\ }\href {https://doi.org/10.1038/nphys2429} {\bibfield  {journal}
  {\bibinfo  {journal} {Nature Physics}\ }\textbf {\bibinfo {volume} {8}},\
  \bibinfo {pages} {795} (\bibinfo {year} {2012})}\BibitemShut {NoStop}%
\bibitem [{\citenamefont {Dom\'{\i}nguez}\ \emph {et~al.}(2017)\citenamefont
  {Dom\'{\i}nguez}, \citenamefont {Kashuba}, \citenamefont {Bocquillon},
  \citenamefont {Wiedenmann}, \citenamefont {Deacon}, \citenamefont {Klapwijk},
  \citenamefont {Platero}, \citenamefont {Molenkamp}, \citenamefont
  {Trauzettel},\ and\ \citenamefont {Hankiewicz}}]{shap3}%
  \BibitemOpen
  \bibfield  {author} {\bibinfo {author} {\bibfnamefont {F.}~\bibnamefont
  {Dom\'{\i}nguez}}, \bibinfo {author} {\bibfnamefont {O.}~\bibnamefont
  {Kashuba}}, \bibinfo {author} {\bibfnamefont {E.}~\bibnamefont {Bocquillon}},
  \bibinfo {author} {\bibfnamefont {J.}~\bibnamefont {Wiedenmann}}, \bibinfo
  {author} {\bibfnamefont {R.~S.}\ \bibnamefont {Deacon}}, \bibinfo {author}
  {\bibfnamefont {T.~M.}\ \bibnamefont {Klapwijk}}, \bibinfo {author}
  {\bibfnamefont {G.}~\bibnamefont {Platero}}, \bibinfo {author} {\bibfnamefont
  {L.~W.}\ \bibnamefont {Molenkamp}}, \bibinfo {author} {\bibfnamefont
  {B.}~\bibnamefont {Trauzettel}},\ and\ \bibinfo {author} {\bibfnamefont
  {E.~M.}\ \bibnamefont {Hankiewicz}},\ }\bibfield  {title} {\bibinfo {title}
  {Josephson junction dynamics in the presence of $2\ensuremath{\pi}$- and
  $4\ensuremath{\pi}$-periodic supercurrents},\ }\href
  {https://doi.org/10.1103/PhysRevB.95.195430} {\bibfield  {journal} {\bibinfo
  {journal} {Phys. Rev. B}\ }\textbf {\bibinfo {volume} {95}},\ \bibinfo
  {pages} {195430} (\bibinfo {year} {2017})}\BibitemShut {NoStop}%
\bibitem [{\citenamefont {Mukherjee}\ \emph {et~al.}(2017)\citenamefont
  {Mukherjee}, \citenamefont {Rao},\ and\ \citenamefont
  {Kundu}}]{mukherjee2017transport}%
  \BibitemOpen
  \bibfield  {author} {\bibinfo {author} {\bibfnamefont {D.~K.}\ \bibnamefont
  {Mukherjee}}, \bibinfo {author} {\bibfnamefont {S.}~\bibnamefont {Rao}},\
  and\ \bibinfo {author} {\bibfnamefont {A.}~\bibnamefont {Kundu}},\ }\bibfield
   {title} {\bibinfo {title} {Transport through andreev bound states in a weyl
  semimetal quantum dot},\ }\href@noop {} {\bibfield  {journal} {\bibinfo
  {journal} {Physical Review B}\ }\textbf {\bibinfo {volume} {96}},\ \bibinfo
  {pages} {161408} (\bibinfo {year} {2017})}\BibitemShut {NoStop}%
\end{thebibliography}%

 \end{document}